\documentclass[reqno,12pt]{amsart}
\usepackage{amssymb, latexsym, euscript}
\newtheorem{theorem}{Theorem}[section]
\newtheorem{proposition}[theorem]{Proposition}

\newtheorem{lemma}[theorem]{Lemma}
\newtheorem{definition}[theorem]{Definition}
\newtheorem{corollary}[theorem]{Corollary}

\theoremstyle{definition}

\newtheorem{remark}[theorem]{Remark}

\numberwithin{equation}{section}

\begin{document}
\title[On $J$-self-adjoint operators with empty resolvent set]{On a class of $J$-self-adjoint operators with empty resolvent set}
\author[S.~Kuzhel]{Sergii~Kuzhel}
\author[C.~Trunk]{Carsten~Trunk}

\address{Institute of Mathematics of the National
Academy of Sciences of Ukraine \\ 01601, Kiev-4,
Ukraine} \email{kuzhel@imath.kiev.ua}

\address{ Institut f\"{u}r Mathematik \\
Technische Universit\"{a}t Ilmenau\\
Postfach 10 05 65, 98684 Ilmenau \\
Germany} \email{carsten.trunk@tu-ilmenau.de}


\maketitle
\begin{abstract}
In the present paper we investigate  the set $\Sigma_J$ of all $J$-self-adjoint extensions of a symmetric operator $S$ with deficiency indices $<2,2>$ which commutes with a non-trivial fundamental symmetry $J$  of a Krein space $(\mathfrak{H}, [\cdot,\cdot])$,
\begin{equation*}
SJ=JS.
\end{equation*}
Our aim is to describe different types of $J$-self-adjoint extensions of $S$.
One of our main results is
the equivalence between the presence of $J$-self-adjoint extensions of $S$ with empty resolvent set
and the commutation of $S$ with a Clifford algebra ${\mathcal C}l_2(J,R)$, where $R$ is an additional fundamental symmetry with $JR=-RJ$.  This enables one
to construct the collection of operators $C_{\chi,\omega}$ realizing the property of stable
$C$-symmetry for extensions $A\in\Sigma_J$ directly in terms of ${\mathcal C}l_2(J,R)$
and to parameterize the corresponding subset  of extensions
 with stable $C$-symmetry. Such a situation occurs naturally in many applications, here we discuss the case of an indefinite Sturm-Liouville operator on the real line and a one dimensional Dirac operator with point interaction.
\end{abstract}

\section{Introduction}
Let $(\mathfrak{H},[\cdot,\cdot])$ be a Krein space with a non-trivial fundamental symmetry $J$ (i.e.,
$J^2=I$, $J\not={\pm{I}}$, and  $(\mathfrak{H},[J\, \cdot,\cdot])$ is a Hilbert space)
and corresponding fundamental decomposition
\begin{equation}\label{Leipzig}
\mathfrak{H}=\mathfrak{H}_+\oplus\mathfrak{H}_-,
\end{equation}
where $\mathfrak{H}_{\pm}=\frac{1}{2}(I{\pm}J)$.
Let $A$ be a linear operator in $\mathfrak{H}$ which is $J$-self-adjoint with
respect to the Krein space inner product $[\cdot,\cdot]$.
In contrast to self-adjoint operators in Hilbert spaces (which
necessarily have a purely real spectrum), $J$-self-adjoint operators $A$,
in general, have spectra $\sigma(A)$ which are only symmetric
with respect to the real axis: $\mu\in\sigma(A)$ if and only if
$\overline{\mu}\in\sigma(A)$.
Moreover, the situation where $\sigma(A)=\mathbb{C}$ (i.e., $A$ has the empty resolvent set)
is also possible.

It is simple to construct
infinitely many $J$-self-adjoint operators with empty resolvent set. For instance, let $\mathcal{K}$ be a Hilbert space and let  $L$ be a symmetric (non-self-adjoint) operator in $\mathcal{K}$. Consider the operators
$$
A:=\left(\begin{array}{cc}
L & 0 \\
0 & L^*
\end{array}
\right), \qquad J=\left(\begin{array}{cc}
0 & I \\
I & 0
\end{array}
\right)
$$
in the product Hilbert space $\mathfrak{H}=\mathcal{K}\oplus\mathcal{K}$. Then $J$ is a fundamental symmetry
in $\mathfrak{H}$ and $A$ is a $J$-self-adjoint operator. As $\rho(L)=\emptyset$, it is clear that $\rho(A)=\emptyset$.

This example shows that the property $\rho(A)=\emptyset$ is a consequence of
the special structure of $A$. It is natural to suppose that this relationship can be made more exact
for some special types of $J$-self-adjoint operators.

In the present paper we investigate such a point by considering the set $\Sigma_J$ of all $J$-self-adjoint extensions $A$ of the symmetric operator $S$ with deficiency indices $<2,2>$ which commutes with $J$:
\begin{equation}\label{es12}
SJ=JS.
\end{equation}
Our aim is to describe different types of $J$-self-adjoint extensions of $S$.
For this let us denote (see Section 2.4 below) by  $\mathfrak{U}$ the set of all fundamental symmetries
which commute with  $S$, by $\Sigma_J^{st}$
we denote the set of all $J$-self-adjoint
extensions of $S$ which commute with a fundamental symmetry in $\mathfrak{U}$, by $\Upsilon_{J}$ the set of all $J$-self-adjoint
extensions of $S$ which commute with $J$ and  by $\Upsilon_{\mathfrak{U}}$, the set of all $J$-self-adjoint
extensions  which commute with all operators in $\mathfrak{U}$.
By definition  we have  $J\in \mathfrak{U}$ and
\begin{equation} \label{Ilmenau}
\Upsilon_{\mathfrak{U}} \subset \Upsilon_{J} \subset \Sigma_J^{st}.
\end{equation}
 Operators from $\Sigma_J^{st}$ are said  to have the property of stable $C$-symmetry, see \cite{KU}. In particular,  they are fundamental reducible and, hence, similar to a self-adjoint operator in a Hilbert space.
$J$-self-adjoint operators with stable $C$-symmetries admit detailed spectral analysis
(like the case of self-adjoint operators), cf.\  \cite{AKG, HK},
and their set $\Sigma_{J}^{st}$ may be used as an exactly solvable model
explaining (at an abstract level) the appearance of exceptional points on the
boundary of the domain of the exact $\mathcal{PT}$-symmetry in ${\mathcal{PT}}$-symmetric quantum mechanics (see \cite{BE, GRS, MO1, MO2} and the references therein).

In the case of a simple symmetric operator $S$, we show in this paper that the existence of at least one $J$-self-adjoint extension of $S$ with empty resolvent set
leads to the quite specific structure of the underlying symmetric operator $S$.
Namely, we have in  \eqref{Ilmenau} strict inclusions,
$$
\Upsilon_{\mathfrak{U}} \subset \Upsilon_{J} \subset \Sigma_J^{st}\quad
(\Upsilon_{\mathfrak{U}} \neq \Upsilon_{J} \neq \Sigma_J^{st}),
$$
which implies a rich structure of extensions with completely different
properties. Moreover, in Corollary \ref{es303} and Theorem \ref{es2020} below we give a full parametrization
of the sets $\Upsilon_{\mathfrak{U}}$, $\Upsilon_{J}$ and $\Sigma_J^{st}$ in terms of (up to) four real parameters.

If, on the other hand, all $J$-self-adjoint extension of $S$ have non-empty
 resolvent set, we show (cf.\ Theorem \ref{es112} below) equality in \eqref{Ilmenau},
 $$
 \Upsilon_{\mathfrak{U}} = \Upsilon_{J} = \Sigma_J^{st}.
 $$
 Moreover, we have $\mathfrak{U}=\{J\}$. This is in particular the case, if there exists at least one definitizable extension (Corollary \ref{es131} below).

We show that
the property of empty resolvent set for a $J$-self-adjoint extension of $S$
is equivalent to one of the following statements.
 \begin{itemize}
 \item There exists  an additional fundamental symmetry $R$ in $\mathfrak{H}$ such that
\begin{equation*}\label{lalalu}
SR=RS, \qquad JR=-RJ.
\end{equation*}
 \item The operator $S_+:=S\upharpoonright_{\mathfrak{H}_+}$ is unitarily equivalent to $S_-:=S\upharpoonright_{\mathfrak{H}_-}$, where
     $\mathfrak{H}_\pm$ are from the fundamental decomposition \eqref{Leipzig} corresponding to $J$.
 \item The  characteristic function $s_+$ of $S_+$ (in the sense of A.\ Straus, see \cite{SH}) is equal (up to the multiplication by an unimodular constant) to the  characteristic function $s_-$ of $S_-$.
 \end{itemize}

If, in addition, the characteristic function $\textsf{Sh}(\cdot)$ of $S$
is not identically equal to zero,
we provide a complete description of the set $\mathfrak{U}$
in terms of $R$ and $J$. More precisely (see Theorem \ref{es35} below),
$\mathfrak{U}$ consists of all operators $C$ of the form
$$
  C=(\cosh\chi)J+(\sinh\chi){JR}[\cos\omega+i(\sin\omega)J]
 $$
with  $\chi\in\mathbb{R}$ and  $\omega\in[0,2\pi)$.

The operators $J$ and $R$ can be interpreted as basis (generating) elements of the complex Clifford
algebra ${\mathcal C}l_2(J,R):=\mbox{span}\{I, J, R, JR\}$  and they give rise to a `rich' family $\Sigma_{J}^{st}$.
 The results of the present paper enables one to claim that the existence of $J$-self-adjoint extensions with empty resolvent set for a symmetric operator $S$ with property (\ref{es12}) and deficiency indices $<2,2>$
 is equivalent to the commutation of $S$ with an arbitrary element of the Clifford algebra ${\mathcal C}l_2(J,R)$.

The paper is structured as follows. Section 2 contains a lot of auxiliary results related to the Krein space theory and the extension theory of symmetric operators. In the latter case we emphasize the usefulness of the Krein spaces ideology for the description of the set $\Sigma_J$ of $J$-self-adjoint extensions of $S$
in terms of unitary $2\times{2}$-matrix $U$ and the definition of the characteristic function $\textsf{Sh}(\cdot)$ of $S$.

In Section 3, we establish a necessary and sufficient condition under which $\Sigma_J$ contains
operators with empty resolvent set (Theorem \ref{es95} and Corollary \ref{es202}) and explicitly describe these operators in terms of unitary matrices $U$ (Corollary \ref{es200}).

In Section 4 we establish our main result (Theorem \ref{es600}) about
the equivalence between the presence of $J$-self-adjoint extensions of $S$ with empty resolvent set
and the commutation of $S$ with a Clifford algebra ${\mathcal C}l_2(J,R)$.  This enables one
to construct the collection of operators $C_{\chi,\omega}$ realizing the property of stable
$C$-symmetry for extensions $A\in\Sigma_J$ directly in terms of ${\mathcal C}l_2(J,R)$ (Theorem \ref{es35})
and to describe the corresponding subset $\Sigma_{J}^{st}$ of extensions
$A\in\Sigma_J$ with stable $C$-symmetry in terms of matrices $U$ (Corollary \ref{es303} and Theorem \ref{es2020}).

Section 5 contains some examples. We consider the case of an indefinite Sturm-Liouville expression on the real line. Then the symmetric operator
$S$ is obtained by imposing additional boundary conditions at zero (which
in some sense decomposes the problem into two differential expressions defined on $\mathbb R_+$ and $\mathbb R_-$, respectively). Then with the results from Section
\ref{Section3} we are able to prove that all $J$-self-adjoint extensions of $S$ have non-empty resolvent set. This extends results from \cite{B07,BP,KarMal}.
Finally, we consider a one dimensional impulse and a Dirac
operator with point perturbation.

Throughout the paper,  the symbols $\mathcal{D}(A)$ and $\mathcal{R}(A)$ denotes the domain and the range of a linear operator $A$. $A\upharpoonright_{\mathcal{D}}$ means the restriction of $A$ onto a
set $\mathcal{D}$. The notation $\sigma(A)$ and $\rho(A)$ are used for the spectrum
and the resolvent set of $A$.
The sign \rule{2mm}{2mm} denotes the end of a proof.

\section{Preliminaries}
\subsection{Elements of the Krein space theory.}
Let $\mathfrak{H}$ be a Hilbert space with inner product
$(\cdot,\cdot)$ and with non-trivial fundamental symmetry $J$ (i.e., $J=J^*$,
$J^2=I$, and $J\not={\pm{I}}$).
The space $\mathfrak{H}$ endowed with the indefinite inner product
(indefinite metric) \ $[\cdot,\cdot]:=(J{\cdot}, \cdot)$ is
called  a  \emph{Krein space}  $(\mathfrak{H}, [\cdot,\cdot])$.
For the basic theory of Krein spaces and operators acting therein we refer  to the monographs \cite{AZ} and \cite{Bog}.

The projectors  $P_{\pm}=\frac{1}{2}(I{\pm}J)$ determine \emph{a fundamental
decomposition} of $\mathfrak{H}$,
\begin{equation}\label{d1}
\mathfrak{H}=\mathfrak{H}_+\oplus\mathfrak{H}_-, \qquad
\mathfrak{H}_-=P_{-}\mathfrak{H}, \quad \mathfrak{H}_+=P_{+}\mathfrak{H},
\end{equation}
where $(\mathfrak{H}_+, [\cdot,\cdot])$ and  $(\mathfrak{H}_-, -[\cdot,\cdot])$ are Hilbert spaces.
With respect to the fundamental decomposition (\ref{d1}), the operator $J$ has the following form
$$
J=\left(\begin{array}{cc}
I & 0 \\
0 & -I
\end{array}
\right).
$$

A subspace $\mathfrak{L}$ of $\mathfrak{H}$ is called \emph{hypermaximal neutral} if
$$
\mathfrak{L}=\mathfrak{L}^{[\bot]}=\{x\in\mathfrak{H}
 :  [x,y]=0, \ \forall{y}\in\mathfrak{L}\}.
$$

A subspace $\mathfrak{L}\subset\mathfrak{H}$ is called {\it uniformly positive
(uniformly negative)} if $[x,x]\geq{a}^2\|x\|^2$ \
(resp.\ $-[x,x]\geq{a}^2\|x\|^2$) $a\in\mathbb{R}$ for all
$x\in\mathfrak{L}$. The subspaces $\mathfrak{H}_{\pm}$ in (\ref{d1}) are
examples of uniformly positive and uniformly negative subspaces and, moreover, they are maximal, i.e.,
$\mathfrak{H}_+$ $(\mathfrak{H}_-)$ is not a proper subspace of an uniformly positive (resp.\ negative) subspace.

Let $\mathfrak{L}_+(\not={\mathfrak H}_+)$ be an arbitrary maximal uniformly positive subspace.
Then its $J$-orthogonal complement $ \mathfrak{L}_-=\mathfrak{L}_+^{[\bot]}$ is
maximal uniformly negative and the direct $J$-orthogonal sum
\begin{equation}\label{d2}
\mathfrak{H}=\mathfrak{L}_+[\dot{+}]\mathfrak{L}_-
\end{equation}
gives a fundamental decomposition of $\mathfrak{H}$.

With respect to (\ref{d2}) we define an operator $C$ via
$$
C=\left(\begin{array}{cc}
I & 0 \\
0 & -I
\end{array}
\right).
$$

We have $C^2=I$ and $C$ is a self-adjoint operator in the Hilbert space $(\mathfrak{H},
(\cdot, \cdot)_C)$, where the inner product $(\cdot, \cdot)_C$ is given by
$$
(x,y)_C:=[Cx,y]=(JCx,y), \quad x,y\in\mathfrak{H}.
$$

Note that $(\cdot,\cdot)_C$ and $(\cdot,\cdot)$ are equivalent, see, e.g., \cite{L3}.
Hence, one can view $C$ as a fundamental symmetry of the Krein space $(\mathfrak{H}, [\cdot,\cdot])$
with an underlying Hilbert space $(\mathfrak{H}, (\cdot,\cdot)_C)$.

Summing up, there is a one-to-one correspondence between the set of all decompositions
(\ref{d2}) of the Krein space $(\mathfrak{H}, [\cdot,\cdot])$ and the set of all bounded
operators $C$ such that
\begin{equation}\label{sos1}
{C}^2=I, \qquad  J{C}>0.
\end{equation}

\begin{definition}\label{dad1}
An operator ${A}$ acting in a Krein space $(\mathfrak{H},
[\cdot,\cdot])$ has the property of  ${C}$-symmetry if
there exists a bounded linear operator ${C}$ in $\mathfrak{H}$ such
that: \ $(i) \ {C}^2=I;$ \quad $(ii) \ {J}C>0$; \quad
$(iii) \ A{C}={C}A$.
\end{definition}

In particular, if $A$ is a ${J}$-self-adjoint operator with the property of $C$-symmetry, then its counterparts
$$
A_\pm:=A\upharpoonright_{{\mathfrak L}_\pm}, \qquad \mathfrak{L}_\pm=\frac{1}{2}(I\pm{C})\mathfrak{H}
$$
are self-adjoint operators in the Hilbert
spaces ${\mathfrak{L}_+}$ and ${\mathfrak{L}_-}$ endowed with the inner
products $[\cdot,\cdot]$ and $-[\cdot,\cdot]$, respectively. This simple
observation leads to the following statement, which is a direct consequence of the Phillips theorem \cite[Chapter 2, Corollary 5.20]{AZ}.
\begin{proposition}\label{sese1}
A ${J}$-self-adjoint operator $A$ has the
property of ${C}$-symmetry if and only if $A$ is similar to a
self-adjoint operator in $\mathfrak{H}$.
\end{proposition}

In conclusion, we emphasize that the notion of $C$-symmetry in Definition \ref{dad1} coincides with the
notion of fundamentally reducible operator (see, e.g., \cite{J1}). However, in the context of this paper and motivated by
\cite{AKG, BE, BK1, GuKu, MO1, MO2}, we prefer to use the notion of $C$-symmetry.

\subsection{Elements of the extension theory in Hilbert spaces.}
Here and in the following we denote by $\mathbb C_+$ ($\mathbb C_-$)
the open upper (resp.\ lower) half plane.
Let $S$ be a closed symmetric densely defined operator with equal deficiency indices acting in the Hilbert space $(\mathfrak{H}, (\cdot,\cdot))$.

We denote by $
\mathfrak{N}_\mu=\ker(S^*-\mu{I})$, $\mu\in\mathbb{C}\setminus\mathbb{R}$,
the defect subspaces of $S$  and consider
the Hilbert space $\mathfrak{M}=\mathfrak{N}_{i}\dot{+}\mathfrak{N}_{-i}$ with
the inner product
\begin{equation}\label{rrr1}
(x,y)_{\mathfrak{M}}=2[(x_i,y_i)+(x_{-i},y_{-i})],
\end{equation}
where $x=x_i+x_{-i}$ and $y=y_i+y_{-i}$ with  $x_{i}, y_{i}\in\mathfrak{N}_i$,
$x_{-i}, y_{-i}\in\mathfrak{N}_{-i}$.

The operator $Z$ which acts as identity
operator $I$ on $\mathfrak{N}_{i}$ and minus identity operator $-I$ on
$\mathfrak{N}_{-i}$ is an example of a fundamental
symmetry in $\mathfrak{M}$.


According to the von-Neumann formulas (see, e.g., \cite{Naimark, KK}) any closed intermediate extension
$A$ of $S$ (i.e.,
$S\subset{A}\subset{S}^*$) in the Hilbert space $(\mathfrak{H}, (\cdot,\cdot))$ is
uniquely determined by the choice of a subspace
$M\subset\mathfrak{M}$:
\begin{equation}\label{e55}
A=S^*\upharpoonright_{\mathcal{D}(A)}, \qquad \mathcal{D}(A)=\mathcal{D}(S)\dot{+}M,
\end{equation}


Let us set $M=\mathfrak{N}_\mu$ ($\mu\in\mathbb{C}_+$) in (\ref{e55}) and denote by
\begin{equation}\label{as908}
A_\mu=S^*\upharpoonright_{\mathcal{D}(A_\mu)}, \qquad \mathcal{D}(A_\mu)=\mathcal{D}(S)\dot{+}\mathfrak{N}_\mu, \quad \forall\mu\in\mathbb{C}_+
\end{equation}
the corresponding maximal dissipative extensions of $S$.
The operator-function
\begin{equation}\label{as909}
\textsf{Sh}(\mu)=(A_\mu-iI)(A_\mu+iI)^{-1}\upharpoonright_{\mathfrak{N}_i} : {\mathfrak{N}_i}\to{\mathfrak{N}_{-i}}, \quad \mu\in\mathbb{C}_+
\end{equation}
is the characteristic function of $S$ defined by A. Straus \cite{SH}.

The characteristic function $\textsf{Sh}(\cdot)$ is connected
with the Weyl function of the symmetric operator $S$ constructed in terms of
boundary triplets (see \cite[p.\ 12]{DM}, \cite[p.\ 1123]{GorKoch}). For instance, if $M(\cdot)$ is the Weyl function
of $S$ associated with the boundary triplet
$(\mathfrak{N}_{i}, \Gamma_0, \Gamma_1)$, where
\begin{equation}\label{sas6}
\Gamma_0{f}=f_{i}+Vf_{-i}, \quad \Gamma_1{f}=if_{i}-iVf_{-i}, \
f=u+f_{i}+f_{-i}\in\mathcal{D}(S^*)
\end{equation}
and $V : \mathfrak{N}_{-i}\to\mathfrak{N}_{i}$ is an arbitrary unitary
mapping, then
\begin{equation}\label{es90}
M(\mu)=i(I+V\textsf{Sh}(\mu))(I-V\textsf{Sh}(\mu))^{-1}, \qquad \mu\in\mathbb{C_+}.
\end{equation}

The function $V\textsf{Sh}(\cdot)$ in (\ref{es90}) coincides with the characteristic function of
$S$ associated with the boundary triplet $(\mathfrak{N}_{i}, \Gamma_0, \Gamma_1)$ \cite{Kochubei}.

Another (equivalent) definition of $\textsf{Sh}(\cdot)$ (see \cite{SH})
is based on the relation
\begin{equation}\label{GDR2}
\mathcal{D}(A_\mu)=\mathcal{D}(S)\dot{+}\mathfrak{N}_\mu=\mathcal{D}(S)\dot{+}(I-\textsf{Sh}(\mu))\mathfrak{N}_i, \quad \mu\in\mathbb{C}_+,
\end{equation}
which also allows one to uniquely determine $\textsf{Sh}(\cdot)$.

The characteristic function $\textsf{Sh}(\cdot)$ can be easily interpreted
in the Krein space setting. Indeed,
according to the von-Neumann formulas, $\mathcal{D}(A_\mu)=\mathcal{D}(S)\dot{+}L_\mu$, where $L_\mu\subset\mathfrak{M}$ is a maximal uniformly positive subspace in the Krein space $(\mathfrak{M}, [\cdot, \cdot]_{Z})$.
Using (\ref{GDR2}), we conclude that $L_\mu=(I-\textsf{Sh}(\mu))\mathfrak{N}_i$ and hence,
$-\textsf{Sh}(\mu)$ is the angular operator of $L_\mu$ with respect to the maximal uniformly positive subspace
$\mathfrak{N}_{i}$ of the Krein space $(\mathfrak{M}, [\cdot, \cdot]_{Z})$
(see \cite{AZ} for the concept of angular operators).

\subsection{Elements of the extension theory in Krein spaces.}

 In what follows we assume that $S$ satisfies (\ref{es12}), where $J$ is a fundamental symmetry in $(\mathfrak{H}, (\cdot,\cdot))$.

The condition (\ref{es12}) immediately leads to the special structure of $S$ with respect to the fundamental decomposition (\ref{d1}):
\begin{equation}\label{ea1b}
S=\left(\begin{array}{cc} S_+  &  0 \\
0 & S_-
\end{array}\right), \qquad S_+=S\upharpoonright_{\mathfrak{H}_+}, \quad S_-=S\upharpoonright_{\mathfrak{H}_-},
\end{equation}
where $S_\pm$ are closed symmetric densely defined operators in $\mathfrak{H}_\pm$.

Denote by $\Sigma_J$ \emph{the collection of all $J$-self-adjoint
extensions of $S$} and set
\begin{equation}\label{les1}
\Upsilon_{J}=\{A\in\Sigma_{J} \ | \
\ AJ=JA \ \}.
\end{equation}
It is clear that $\Upsilon_{J}\subset\Sigma_{J}$ and
an arbitrary $A\in\Upsilon_{J}$ is, simultaneously, self-adjoint and
$J$-self-adjoint extensions of $S$. The set
$\Upsilon_{J}$ \emph{is non-empty if and only if each symmetric operator $S_{\pm}$ in}
(\ref{ea1b})  \emph{has equal deficiency indices}.
We always suppose that $\Upsilon_{J}\not=\emptyset$.

Since $S$ satisfies (\ref{es12}) \emph{the
subspaces $\mathfrak{N}_{\pm{i}}$ reduce $J$} and the restriction
$J\upharpoonright\mathfrak{M}$ gives rise to a fundamental
symmetry in the Hilbert space $\mathfrak{M}$. Moreover, according to
the properties of $Z$ mentioned above, $JZ=ZJ$ and $JZ$ is
 a fundamental symmetry in $\mathfrak{M}$. Therefore, the
sesquilinear form
\begin{equation}\label{aaaa1}
[x,y]_{JZ}=(JZx,y)_{\mathfrak{M}}=2[(Jx_i,y_i)-(Jx_{-i},y_{-i})]
\end{equation}
defines an indefinite metric on $\mathfrak{M}$.

It is known (see, e.g., \cite[Proposition 3.1]{AKG}) that  an arbitrary $J$-self-adjoint extension $A$ of $S$ is uniquely determined by (\ref{e55}), where $M$ is a \emph{hypermaximal neutral subspace} of the Krein space $(\mathfrak{M}, [\cdot, \cdot]_{JZ})$.

In comparison with self-adjoint extensions in the sense of Hilbert spaces, we remark that \emph{self-adjoint extensions} of $S$ in $(\mathfrak{H}, (\cdot,\cdot))$ are also determined by {(\ref{e55})} but then subspaces $M$
are assumed to be hypermaximal neutral in the Krein space $(\mathfrak{M}, [\cdot, \cdot]_{Z})$ with the indefinite metric (cf.\ (\ref{aaaa1}))
$$
[x,y]_{Z}=(Zx,y)_{\mathfrak{M}}=2[(x_i,y_i)-(x_{-i},y_{-i})].
$$

\subsection{$J$-self-adjoint operators with stable ${C}$-symmetries}
Denote by $\mathfrak{U}$ the set of \emph{all
possible $C$-symmetries} of the symmetric operator $S$.
By Definition \ref{dad1}, this means that
$$
C\in\mathfrak{U} \quad \iff \quad {C}^2=I, \quad JC>0, \quad SC=CS.
$$

The next result directly follows from \cite{AKG}. We  repeat principal stages for the reader's convenience.
\begin{lemma}\label{aaaa2}
The set $\mathfrak{U}$ is non-empty and $C\in\mathfrak{U}$ if and only if $C^*\in\mathfrak{U}$.
\end{lemma}
\emph{Proof.}
It follows from (\ref{es12}) that ${J}\in\mathfrak{U}$. Therefore,  $\mathfrak{U}\not=\emptyset$.

Let $C\in\mathfrak{U}$. The conditions ${C}^2=I$ and $JC>0$ are equivalent to the presentation $C=Je^Y$, where $Y$ is a bounded
self-adjoint operator in $\mathfrak{H}$ such that $JY=-YJ$ \cite[Remark 2.1]{AKG}.
In that case $C^*=Je^{-Y}$ and, obviously, $C^*$ satisfies the relations ${C^*}^2=I $ and $JC^*>0$.

Since $S$ commutes with $J$ and $C$ one gets $Se^Y=e^Y{S}$. But then $SC^*=Se^Y{J}=e^Y{J}S=C^*S.$
Hence, $C^*\in\mathfrak{U}$.
\rule{2mm}{2mm}
\begin{definition}[\cite{KU}]\label{dad34}
An operator $A\in\Sigma_J$ has the property of stable $C$-symmetry if
$A$ and $S$ have the property of $C$-symmetry realized by the \emph{same} operator
$C$, i.e., there exists $C\in\mathfrak{U}$ with $AC=CA$.  \end{definition}

Denote
\begin{equation}\label{les9}
\Sigma_{J}^{st}=\{A\in\Sigma_{J} \ | \ \exists C\in\mathfrak{U} \ \mbox{such that}
\ AC=CA\}.
\end{equation}

Due to Definition \ref{dad34}, $\Sigma_{J}^{st}$ consists of $J$-self-adjoint extensions $A$ of $S$ with the property of stable $C$-symmetry. It follows from (\ref{les1}) and (\ref{les9}) that
$\Sigma_{J}^{st}\supset\Upsilon_{J}$. Hence, $\Sigma_{J}^{st}$ is non-empty. 

Denote
\begin{equation}
\Upsilon_{\mathfrak{U}}=\{A\in\Sigma_{J} \ | \
\ AC=CA, \ \forall{C}\in\mathfrak{U}\}.
\end{equation}
It is clear that
\begin{equation}\label{les6}
\Upsilon_{\mathfrak{U}}\subset\Upsilon_{J}\subset\Sigma_{J}^{st}\subset\Sigma_{J}.
\end{equation}

The next Theorem gives a condition for the non-emptiness
of the left-hand side of the chain (\ref{les6}).

\begin{theorem}\label{sssr10}
If the characteristic function $\textsf{Sh}(\cdot)$ of $S$ is boundedly invertible for at least one
$\mu\in\mathbb{C}_+$, then $\Upsilon_{\mathfrak{U}}\not=\emptyset$.
\end{theorem}

\emph{Proof.} Let $C\in\mathfrak{U}$. Then $S^*C=CS^*$  (see the proof of Lemma \ref{aaaa2}) and, hence,
\begin{equation}\label{es404}
C : \mathfrak{N}_\mu{\to}\mathfrak{N}_\mu,  \qquad \forall\mu\in\mathbb{C}\setminus\mathbb{R}.
\end{equation}
Therefore, $A_\mu{C}=CA_\mu$ for
maximal dissipative extensions $A_\mu$ of $S$ (see (\ref{as908})).
This means that the characteristic function $\textsf{Sh}(\cdot)$ defined by (\ref{as909})
commutes with an arbitrary $C\in\mathfrak{U}$, i.e.,
\begin{equation}\label{es407}
\textsf{Sh}(\mu)C=C\textsf{Sh}(\mu), \qquad \forall\mu\in\mathbb{C}_+, \quad \forall{C}\in\mathfrak{U}.
\end{equation}

It follows from Lemma \ref{aaaa2} and (\ref{es407}) that
$\textsf{Sh}(\mu)C^*=C^*\textsf{Sh}(\mu)$. Therefore,
\begin{equation}\label{aaaa3}
\textsf{Sh}^*(\mu)C=C\textsf{Sh}^*(\mu), \qquad \forall\mu\in\mathbb{C}_+, \quad \forall{C}\in\mathfrak{U}.
\end{equation}

Let $\textsf{Sh}(\mu)$ be boundedly invertible for a certain $\mu\in\mathbb{C_+}$ and
let $V:\mathfrak{N}_{i}\to\mathfrak{N}_{-i}$  be the isometric factor in the polar decomposition of  $\textsf{Sh}(\mu)$. Then $VC=CV$ for all $C\in\mathfrak{U}$ (since (\ref{es407}) and (\ref{aaaa3})).
This means that the operator
$$
A=S^*\upharpoonright_{\mathcal{D}(A)}, \qquad \mathcal{D}(A)=\mathcal{D}(S)\dot{+}\{(I+V)\mathfrak{N}_{i}\}
$$
belongs to $\Upsilon_{\mathfrak{U}}$. \rule{2mm}{2mm}

\begin{remark} The similar result was established by Kochubei \cite[Theorem 1]{KO} for the collection
of unitary operators $\mathfrak{U}=\{U\}$ with the property that $U\in\mathfrak{U}$ implies
$U^{*}\in\mathfrak{U}$.
\end{remark}

According to (\ref{es404}), an arbitrary ${C}\in\mathfrak{U}$ determines
two operators $C\upharpoonright_{\mathfrak{N}_{\pm{i}}}$ acting in ${\mathfrak{N}_{\pm{i}}}$.
\begin{lemma}\label{neww41}
If $S$ is a simple symmetric operator, then the correspondence
${{C}\in\mathfrak{U}}\rightarrow\{C\upharpoonright_{\mathfrak{N}_{{i}}}, C\upharpoonright_{\mathfrak{N}_{{-i}}}\}$ is injective.
\end{lemma}
\emph{Proof.}
Assume the existence of an operator pair $\{C\upharpoonright_{\mathfrak{N}_{i}}, C\upharpoonright_{\mathfrak{N}_{-i}}\}$ for two different operators $C, \ \widetilde{C}\in\mathfrak{U}$.
Then $(C-\widetilde{C})\mathcal{D}(S^*)\subset\mathcal{D}(S)$.
Therefore, $(C-\widetilde{C})\mathfrak{N}_\mu\subset\mathcal{D}(S)$.
On the other hand, $(C-\widetilde{C})\mathfrak{N}_\mu\subset\mathfrak{N}_\mu$ by (\ref{es404}). The obtained
relations yield $Cf_\mu=\widetilde{C}f_\mu$ for any $f_\mu\in\mathfrak{N}_\mu$ and $\mu\in\mathbb{C}\setminus\mathbb{R}$.
This means that $C=\widetilde{C}$ (since the symmetric operator $S$ is simple). \rule{2mm}{2mm}

\section{Necessary and sufficient condition under which $\Sigma_J$ contains elements with empty resolvent set.}\label{Section3}
In what follows we assume that the deficiency indices of operators $S_\pm$ in (\ref{ea1b}) is $<1,1>$.
In that case, the defect subspaces $\mathfrak{N}_{\pm{i}}(S_\pm)$ of $S_\pm$ are
one-dimensional and
$$
\begin{array}{lr}
\mathfrak{N}_{i}(S_+)=(I+Z)(I+J)\mathfrak{M}; \ & \ \mathfrak{N}_{-i}(S_+)=(I-Z)(I+J)\mathfrak{M}; \vspace{5mm} \\
\mathfrak{N}_{i}(S_-)=(I+Z)(I-J)\mathfrak{M}; \ & \ \mathfrak{N}_{-i}(S_-)=(I-Z)(I-J)\mathfrak{M}.
\end{array}
$$
Hence,  $\mathfrak{N}_{\pm{i}}(S_\pm)$ are orthogonal in the Hilbert space $(\mathfrak{M}, (\cdot,\cdot)_{\mathfrak{M}})$ (see (\ref{rrr1})).

Let $\{e_{++}, e_{+-}, e_{-+}, e_{--}\}$ be an orthogonal basis of $\mathfrak{M}$ such
that
\begin{equation}\label{ura1}
\begin{array}{c}
\mathfrak{N}_{i}(S_{+})=\ker(S^*_+-iI)=\mbox{span}\{e_{++}\},\\[2mm]
\mathfrak{N}_{i}(S_-)=\ker(S^*_--{i}I)=\mbox{span}\{e_{+-}\},\\[2mm]
\mathfrak{N}_{-i}(S_+)=\ker(S^*_+ +iI)=\mbox{span}\{e_{-+}\},\\[2mm]
\mathfrak{N}_{-i}(S_{-})=\ker(S^*_-+iI)=\mbox{span}\{e_{--}\},
\end{array}
\end{equation}
and the elements $e_{++}, e_{+-}, e_{-+}, e_{--}$ have \emph{equal} norms in $\mathfrak{M}$.
It follows from the definition of $e_{\pm\pm}$ that
\begin{equation}\label{ss1}
\begin{array}{c}
Ze_{++}=e_{++}, \ \ Ze_{+-}=e_{+-}, \ \ Ze_{-+}=-e_{-+}, \ \ Ze_{--}=-e_{--} \\[2mm]
Je_{++}=e_{++}, \ \ Je_{+-}=-e_{+-}, \ \ Je_{-+}=e_{-+}, \ \ Je_{--}=-e_{--}
\end{array}
\end{equation}

Relations (\ref{ss1}) mean that the fundamental decomposition of the Krein space
$(\mathfrak{M}, [\cdot, \cdot]_{JZ})$ has the form
\begin{equation}\label{e5}
\mathfrak{M}={\mathfrak{M}}_{-}\oplus{\mathfrak{M}}_{+}, \qquad
{\mathfrak{M}}_{-}=\mbox{span}\{e_{+-}, e_{-+}\}, \quad
{\mathfrak{M}}_{+}=\mbox{span}\{e_{++}, e_{--}\}.
\end{equation}

According to the general theory of Krein spaces \cite[Chapter 1, Theorem 8.10]{AZ}, an arbitrary hypermaximal
neutral subspace $M$ of $(\mathfrak{M}, [\cdot, \cdot]_{JZ})$ is
uniquely determined by an unitary mapping of ${\mathfrak{M}}_{-}$
onto ${\mathfrak{M}}_{+}$. Since
$\dim{\mathfrak{M}}_{\pm}=2$ the set of unitary
mappings ${\mathfrak{M}}_{-}\to{\mathfrak{M}}_{+}$ is in one-to-one correspondence with
the set of unitary matrices
\begin{equation}\label{eee6}
U=e^{i\phi}\left(\begin{array}{cc} qe^{i\gamma} & re^{i\xi} \\
-re^{-i\xi} & qe^{-i\gamma} \end{array}\right), \quad q^2+r^2=1, \quad q, r\in\mathbb{R}_+,
\quad \phi, \gamma, \xi\in{[0,2\pi)}.
\end{equation}

In other words, formulas (\ref{e5}),
(\ref{eee6}) allow one to describe a hypermaximal neutral subspace
$M$ of $(\mathfrak{M}, [\cdot, \cdot]_{JZ})$ as a linear span
\begin{equation}\label{e6}
M=\mbox{span}\{d_1,d_2\}
\end{equation}
of elements
\begin{equation}\label{aaa5}
\begin{array}{c}
d_1=e_{++}+qe^{i(\phi+\gamma)}e_{+-}+re^{i(\phi+\xi)}e_{-+}; \\
d_2=e_{--}-re^{i(\phi-\xi)}e_{+-}+qe^{i(\phi-\gamma)}e_{-+}.
\end{array}
\end{equation}

This means that (\ref{eee6}) - (\ref{aaa5}) establish a one-to-one
correspondence between domains
$\mathcal{D}(A)=\mathcal{D}(S)\dot{+}M$ of
$J$-self-adjoint extensions $A$ of $S$ and
unitary matrices $U$. To underline this relationship \emph{we
will use the notation $A_{U}$ for the corresponding $J$-self-adjoint
extension $A$}.

It follows from (\ref{es407}) (with $C=J$) that the characteristic
function $\textsf{Sh}(\cdot):\mathfrak{N}_i\to\mathfrak{N}_{-i}$ commutes with $J$.
Combining this fact with the obvious presentations
\begin{equation}\label{GDR5}
\begin{array}{c}
{\mathfrak{N}}_{i}={\mathfrak{N}}_{i}(S_+)\oplus{\mathfrak{N}}_{i}(S_-)=\mbox{span}\{e_{++}, e_{+-}\}, \vspace{3mm} \\
{\mathfrak{N}}_{-i}={\mathfrak{N}}_{-i}(S_+)\oplus{\mathfrak{N}}_{-i}(S_-)=\mbox{span}\{e_{-+}, e_{--}\}
\end{array}
\end{equation}
and relations (\ref{GDR2}), (\ref{ss1}), we arrive at the conclusion that
\begin{equation}\label{GDR1}
\textsf{Sh}(\mu)e_{++}=s_+(\mu)e_{-+}, \qquad \textsf{Sh}(\mu)e_{+-}=s_-(\mu)e_{--},
\end{equation}
where $s_j$ are holomorphic functions in $\mathbb{C}_+$.
Moreover, it is easy to see that relations in (\ref{GDR1}) determine
the characteristic functions
\begin{equation}\label{GDR3}
\textsf{Sh}_+(\mu):\mathfrak{N}_i(S_+)\to\mathfrak{N}_{-i}(S_+), \quad \textsf{Sh}_-(\mu):\mathfrak{N}_i(S_-)\to\mathfrak{N}_{-i}(S_-)
\end{equation}
of the symmetric operators $S_+$ and $S_-$, respectively.

We will use the notation
$$
s_+\approx{s}_-
$$
if the identity
$e^{i\alpha}{s}_+(\mu)={s}_-(\mu)$ holds for
 for all $\mu\in\mathbb{C}_+$ and for a certain choice of an unimodular constant $e^{i\alpha}$, i.e., the sign $\approx$ means the equality up to the multiplication by an unimodular constant.

\begin{theorem}\label{es95}
Assume that the deficiency indices of operators $S_\pm$ in the presentation (\ref{ea1b}) of $S$ are $<1,1>$.
Then $J$-self-adjoint extensions of $S$ with empty resolvent set exist
if and only if ${s}_+\approx{s}_-$.
\end{theorem}
\emph{Proof.}
It follows from (\ref{GDR2}) that a $J$-self-adjoint extension $A_{U}$ of $S$  with the domain
$\mathcal{D}(A_{U})=\mathcal{D}(S)\dot{+}M$ has a non-real eigenvalue $\mu\in\mathbb{C}_+$
if and only if $U$ has a nontrivial intersection with the subspace $L_\mu=(I-\textsf{Sh}(\mu))\mathfrak{N}_i$.
Therefore,
$$
\sigma(A_{U})\supset\mathbb{C}_+ \quad \mbox{if and only if} \quad {M{\cap}L_\mu\not=\{0\}} \quad \forall{\mu}\in\mathbb{C}_+.
$$

Since $A_{U}$ is a $J$-self-adjoint operator, the inclusion $\sigma(A_{U})\supset\mathbb{C}_+$
is equivalent to $\sigma(A_{U})=\mathbb{C}$.

In view of (\ref{GDR5}) and (\ref{GDR1}), $L_\mu=(I-\textsf{Sh}(\mu))\mathfrak{N}_i=\mbox{span}\{c_1(\mu), c_2(\mu)\}$, where
\begin{equation}\label{es103}
c_1(\mu)=e_{++}-s_+(\mu)e_{-+}, \qquad c_2(\mu)=e_{+-}-s_-(\mu)e_{--}.
\end{equation}
Therefore, the relation ${M{\cap}L_\mu\not=\{0\}}$ holds if and only if  the equation
\begin{equation}\label{es102}
x_1d_1+x_2d_2=y_1c_1(\mu)+y_2c_2(\mu)
\end{equation}
has a nontrivial solution $x_1,x_2,y_1,y_2\in\mathbb{C}$ for all ${\mu}\in\mathbb{C}_+$.
Substituting (\ref{aaa5}) and (\ref{es103}) into (\ref{es102}) and combining the corresponding coefficients
for $e_{\pm\pm}$ we obtain four relations
$$
\begin{array}{lr}
x_1=y_1, &  x_1qe^{i(\phi+\gamma)}-x_2re^{i(\phi-\xi)}=y_2 \\
x_2=-y_2s_-(\mu), & x_1re^{i(\phi+\xi)}+x_2qe^{i(\phi-\gamma)}=-s_+(\mu)y_1
\end{array}
$$
or
$$
\begin{array}{l}
qe^{i(\phi+\gamma)}y_1-(1-re^{i(\phi-\xi)}s_-(\mu))y_2=0, \vspace{2mm} \\
(re^{i(\phi+\xi)}+s_+(\mu))y_1-qe^{i(\phi-\gamma)}s_-(\mu)y_2=0.
\end{array}
$$
The last system has a nontrivial solution $y_1,y_2$ for all ${\mu}\in\mathbb{C}_+$
if and only if its determinant
$$
\left|\begin{array}{lr}
qe^{i(\phi+\gamma)} & -1+re^{i(\phi-\xi)}s_-(\mu) \\
re^{i(\phi+\xi)}+s_+(\mu) & -qe^{i(\phi-\gamma)}s_-(\mu)
\end{array}
\right|=0, \quad \forall{\mu}\in\mathbb{C}_+.
$$
This is the case if and only if
\begin{equation}\label{es104}
e^{2i\phi}s_-(\mu)=re^{i(\phi+\xi)}+s_+(\mu)-re^{i(\phi-\xi)}s_-(\mu)s_+(\mu), \quad \forall{\mu}\in\mathbb{C}_+
\end{equation}

Further, $\textsf{Sh}(i)=0$ by the construction (see (\ref{as909}) or (\ref{GDR2})).
Hence $s_+(i)=s_-(i)=0$ and relation (\ref{es104}) takes the form
$re^{i(\phi+\xi)}=0$ (for $\mu=i$) which means that ${r=0}$. Therefore,
an operator $A_{U}\in\Sigma_J$ has empty resolvent set if and only
\begin{equation}\label{es565}
e^{2i\phi}s_-(\mu)=s_+(\mu),  \qquad \forall{\mu}\in\mathbb{C}_+.
\end{equation}
\rule{2mm}{2mm}

\begin{corollary}\label{es200}
If ${s}_+\approx{s}_-$, then operators $A_{U}\in\Sigma_J$ with empty resolvent set
are determined by the matrices:
\begin{equation}\label{es400}
U=e^{i\phi}\left(\begin{array}{cc}
e^{i\gamma} & 0 \\
0 & e^{-i\gamma} \end{array}\right), \qquad \gamma\in[0,2\pi),
\end{equation}
where $\phi\in[0,2\pi)$ is uniquely determined by (\ref{es565}) if $\textsf{Sh}\not\equiv{0}$ and $\phi$
 is an arbitrary parameter if $\textsf{Sh}\equiv{0}$.
\end{corollary}

\begin{corollary}\label{es202}
Let $S$ be a simple symmetric operator. Then $\Sigma_{J}$ contains operators with empty resolvent set if and only if the operators $S_\pm$ in
(\ref{ea1b}) are unitarily equivalent.
\end{corollary}
\emph{Proof.}
 Assume that $\Sigma_{J}$ contains operators with empty resolvent set and $\textsf{Sh}\not\equiv{0}$.
 Then ${s}_+\not\equiv{0}$ and (\ref{es565}) holds for a certain $\phi\in[0,2\pi)$.
Consider unitary mappings $V_{\pm}:\mathfrak{N}_{-i}(S_\pm)\to\mathfrak{N}_{i}(S_\pm)$
defined by the relations
$$
V_+e_{-+}=e_{++}, \qquad V_-e_{--}=e^{2i\phi}e_{+-}.
$$
By virtue of  (\ref{GDR1}) and (\ref{GDR3}), we get
\begin{equation}\label{es304}
\begin{array}{l}
V_+\textsf{Sh}_+(\mu)e_{++}=s_+(\mu)e_{++}, \vspace{3mm}  \\
V_-\textsf{Sh}_-(\mu)e_{+-}=e^{2i\phi}s_-(\mu)e_{+-}=s_+(\mu)e_{+-}.
\end{array}
\end{equation}
 Then $V_+\textsf{Sh}_+(\cdot)$ and $V_-\textsf{Sh}_-(\cdot)$ are the characteristic
functions (in the sense of \cite{Kochubei})  of $S_\pm$ associated with the boundary triplets $(\mathfrak{N}_i(S_\pm), \Gamma_0, \Gamma_1)$ of $S_{\pm}^*$  defined by (\ref{sas6}). Identifying the defect subspaces
$\mathfrak{N}_i(S_+)=\mbox{span}\{e_{++}\}$  and $\mathfrak{N}_i(S_-)=\mbox{span}\{e_{+-}\}$ with $\mathbb{C}$
and using (\ref{es304})
we arrive at the conclusion that the characteristic
functions of $S_\pm$ associated with boundary triplets $(\mathbb{C}, \Gamma_0, \Gamma_1)$ coincide.

The same is true when $\textsf{Sh}\equiv{0}$. In that case, ${s}_+\equiv{s}_-\equiv{0}$ and the characteristic functions $\textsf{Sh}_\pm$ of $S_\pm$ are equal to zero.

Since $S$ is a simple symmetric operator, $S_\pm$ are also simple symmetric
operators. In that case, the equality of characteristic functions of $S_\pm$ implies
the unitary equivalence of $S_{\pm}$ \cite{GorKoch, Kochubei}.

Conversely, if $S_\pm$ are unitarily equivalent then $S_+=W^{-1}S_-W$, where $W$ is
an unitary mapping of $\mathfrak{H}_+$ onto $\mathfrak{H}_-$.
Therefore,
\begin{equation}\label{es190}
W:\mathfrak{N}_\mu(S_+)\to\mathfrak{N}_\mu(S_-) \quad \mbox{and} \quad W\textsf{Sh}_+(\mu)=\textsf{Sh}_-(\mu)W.
\end{equation}
Assuming $\mu=\pm{i}$ in the first identity of (\ref{es190}) and using (\ref{GDR5}), we find $w_1, w_2\in\mathbb{C}$ with
\begin{equation}\label{es78}
We_{++}=w_1e_{+-}, \quad We_{-+}=w_2e_{--}, \quad |w_1|=|w_2|=1.
\end{equation}

It follows from (\ref{GDR1}) and (\ref{es78}) that
 $W\textsf{Sh}_+(\mu)e_{++}=s_+(\mu)We_{-+}=w_2s_+(\mu)e_{--}$ \ and \
$\textsf{Sh}_-(\mu)We_{++}=w_1\textsf{Sh}_-(\mu)e_{+-}=w_1s_-(\mu)e_{--}$.
Combining the last two identities with the second relation in (\ref{es190}) ones get
$e^{2i\phi}s_-(\mu)=s_+(\mu)$, where $e^{2i\phi}=w_1/w_2$. The statement of
Corollary \ref{es202} follows now from Theorem \ref{es95}.
\rule{2mm}{2mm}

\section{$J$-self-adjoint extensions with empty resolvent set}
As above the deficiency indices of operators $S_\pm$ in the presentation (\ref{ea1b}) of
$S$ are supposed to be $<1,1>$.
In the following we discuss the different situations which can occur:
\begin{itemize}
  \item no member of $\Sigma_J$ has non-empty resolvent set; \vspace{2mm}
  \item there are members of $\Sigma_J$ with empty resolvent set, where we discuss the cases
$\textsf{Sh}(\cdot)\not\equiv{0}$ and $\textsf{Sh}(\cdot)\equiv{0}$, separately (cf.\ Sections 4.2.1 and 4.2.2).
\end{itemize}

\subsection{The set $\Sigma_J$ has no operators with empty resolvent set.}
\begin{theorem}\label{es112}
If $\Sigma_J$ has no operators with empty resolvent set, then
$$\Upsilon_{\mathfrak{U}}=\Upsilon_{J}=\Sigma_{J}^{st}$$ in (\ref{les6}). Moreover, if $S$ is a simple symmetric operator, then
$\mathfrak{U}=\{J\}$.
\end{theorem}

\emph{Proof.}
Let ${C}\in\mathfrak{U}$. It follows from (\ref{es404}) that the operator
$C\upharpoonright_{\mathfrak{N}_{\pm{i}}}$  acts in ${\mathfrak{N}_{\pm{i}}}$ and satisfies
the relations
\begin{equation}\label{es405}
(C\upharpoonright_{\mathfrak{N}_{\pm{i}}})^2=I, \qquad JC\upharpoonright_{\mathfrak{N}_{\pm{i}}}>0.
\end{equation}

Denote by $\mathcal{C}_1$ and $\mathcal{C}_2$
the $2\times{2}$-matrix representations of $C\upharpoonright_{\mathfrak{N}_{{i}}}$ and
$C\upharpoonright_{\mathfrak{N}_{{-i}}}$ with respect to the
orthogonal bases $e_{++}, e_{+-}$  and $e_{-+}, e_{--}$ of $\mathfrak{N}_{{i}}$ and $\mathfrak{N}_{{-i}}$,
respectively.  Then (\ref{es405}) takes the form
\begin{equation}\label{es123}
\mathcal{C}_j^2=\left(\begin{array}{cc} 1  &  0 \\
0 & 1
\end{array}\right), \qquad  \left(\begin{array}{cc} 1  &  0 \\
0 & -1
\end{array}\right)\mathcal{C}_j>0,  \quad j=1,2
\end{equation}
(since $J\upharpoonright_{\mathfrak{N}_{\pm{i}}}$ are determined by (\ref{ss1})).
The Hermiticity of the matrix in the second relation of (\ref{es123}) enables one to deduce
that a matrix $\mathcal{C}_j$ satisfy (\ref{es123}) if and
only if
\begin{equation}\label{es124}
\mathcal{C}_j=\mathcal{C}_{\chi_j,\omega_j}:=\left(\begin{array}{cc}
\cosh\chi_j & (\sinh\chi_j)e^{-i\omega_j} \\
-(\sinh\chi_j)e^{i\omega_j} & -\cosh\chi_j
\end{array}\right), \quad  \chi_j\in\mathbb{R}, \ \omega_j\in[0,2\pi).
\end{equation}

Combining (\ref{es407}) with (\ref{GDR1}) and (\ref{es124}) we get
\begin{eqnarray}\label{es408}
\displaystyle{\left(\begin{array}{cc}
s_+(\mu) & 0 \\
0 & s_-(\mu)
\end{array}\right)\left(\begin{array}{cc}
\cosh\chi_1 & (\sinh\chi_1)e^{-i\omega_1} \\
-(\sinh\chi_1)e^{i\omega_1} & -\cosh\chi_1
\end{array}\right)} & & \\
\displaystyle{=\left(\begin{array}{cc}
\cosh\chi_2 & (\sinh\chi_2)e^{-i\omega_2} \\
-(\sinh\chi_2)e^{i\omega_2} & -\cosh\chi_2
\end{array}\right)\left(\begin{array}{cc}
s_+(\mu) & 0 \\
0 & s_-(\mu)
\end{array}\right)} & & \nonumber
\end{eqnarray}
for matrix representations $\mathcal{C}_{\chi_j,\omega_j}$ of the operators $C\upharpoonright_{\mathfrak{N}_{\pm{i}}}$.

If $\Sigma_J$ has no operators with empty resolvent set, then ${s}_+\not\approx{s}_-$
(Theorem \ref{es95}).  In that case identity (\ref{es408}) holds only in the case $\chi_1=\chi_2=0$, i.e.,
$
\mathcal{C}_{0,\omega_1}=\mathcal{C}_{0,\omega_2}=\left(\begin{array}{cc}
1 & 0 \\
0 & -1
\end{array}\right)$. Therefore, if ${s}_+\not\approx{s}_-$ then
\begin{equation}\label{es501}
C\upharpoonright_{\mathfrak{N}_{\pm{i}}}=J\upharpoonright_{\mathfrak{N}_{\pm{i}}}, \qquad
\forall{C}\in\mathfrak{U}.
\end{equation}

Let us consider an arbitrary $A_U\in\Sigma_{J}^{st}$.
Then  $A_{U}C=CA_{U}$ for  some choice of $C\in\mathfrak{U}$.
It is known that $A_{U}C=CA_{U}$ if and only if $CM=M$, where $M$ is defined by (\ref{e6}) and (\ref{aaa5}) \cite[Theorem 3.1]{AKG}.
This and (\ref{es501}) give
$CM=M$ if and only if $JM=M$.  Therefore, $A_{U}J=JA_{U}$ and $A_{U}\in\Upsilon_J$.
Thus $\Upsilon_J=\Sigma_{J}^{st}$. The identity $\Upsilon_{\mathfrak{U}}=\Upsilon_J$ is verified in the similar manner.

If $S$ is a simple symmetric operator, then $\mathfrak{U}=\{J\}$ due to Lemma \ref{neww41} and relation
(\ref{es501}). \rule{2mm}{2mm}

Recall, that a $J$-selfadjoint operator $A$ in a Krein space
$(\mathfrak{H},[\cdot,\cdot])$ is called
definitizable  (see \cite{L3}) if
$\rho(A) \ne \emptyset$ and there exists a
rational function $p \ne 0$ having poles only in $\rho(A)$
such that $[p(A)x,x] \geq 0$ for all $x \in {\mathfrak{H}}$.

\begin{corollary}\label{es131}
If $\Sigma_{J}$ contains at least one definitizable operator, then
$\Upsilon_{\mathfrak{U}}=\Upsilon_{J}=\Sigma_{J}^{st}$.
\end{corollary}

\emph{Proof.} If $A\in\Sigma_{J}$ is definitizable then an arbitrary operator from
$\Sigma_{J}$ is also definitizable \cite{ABT}. Therefore, $\Sigma_{J}$ has no operators with
empty resolvent sets. \rule{2mm}{2mm}

\subsection{The set $\Sigma_J$ contains operators with empty resolvent set.} $\mbox{}$ \\

In that case two quite different arrangements for the sets
$\Upsilon_{\mathfrak{U}}$, $\Upsilon_{J}$, and $\Sigma_{J}^{st}$
are possible and they will be discussed in Sections 4.2.1 and 4.2.2
below.

We recall
that  $\Sigma_J$ contains operators with empty resolvent set if and only if
$e^{i\alpha}{s}_+(\mu)={s}_-(\mu)$, $\mu \in \mathbb C_+$,
 for a certain parameter $e^{i\alpha}$
(Theorem \ref{es95}). Here, the functions ${s}_\pm(\cdot)$ are defined
in (\ref{GDR1}) with the use of elements $\{e_{\pm\pm}\}$ which are determined up to
the multiplication with an unimodular constants. Therefore, without loss of generality,
we may assume
\begin{equation}\label{es498}
{s}_+={s}_-.
\end{equation}
\begin{theorem}\label{es600}
Let $S$ be a simple symmetric operator. Then the set $\Sigma_J$ contains operators with empty resolvent set if and only if
there exists a fundamental symmetry $R$ (i.e., $R^2=I$ and $R=R^*$) in $\mathfrak{H}$
such that
  \begin{equation}\label{es601}
  SR=RS, \qquad JR=-RJ.
  \end{equation}
\end{theorem}
\emph{Proof.} By virtue of Corollary \ref{es202}, the existence of $J$-self-adjoint extensions of $S$ with empty resolvent set implies that
the symmetric operators $S_\pm$ in (\ref{ea1b}) are unitarily equivalent.
Hence, $S_+=W^{-1}S_-W$, where $W$ is an isometric mapping of $\mathfrak{H}_+$ onto $\mathfrak{H}_-$. It is clear that the operator
\begin{equation}\label{es603}
R=\left(\begin{array}{cc} 0 & W^{-1} \\
W & 0 \end{array}\right)
\end{equation}
determined with respect to the fundamental decomposition (\ref{d1}) is a fundamental symmetry in
$\mathfrak{H}$ and satisfies (\ref{es601}).

Conversely, if (\ref{es601}) holds, then
$S_+=RS_-R.$ Therefore $S_\pm$ are unitarily equivalent and $\Sigma_J$ contains elements with empty resolvent
set (Corollary \ref{es202}). \rule{2mm}{2mm}
\begin{remark}\label{did234}
We do not need the condition of \emph{simplicity} of $S$ in Theorem \ref{es600} if
relations (\ref{es601}) hold. Indeed, $\mathfrak{H}_1=\bigcap_{\mu\in\mathbb{C}\setminus\mathbb{R}}\mathcal{R}(S-\mu{I})$ is the maximal
subspace invariant for $S$ on which the operator $S_1=S\upharpoonright_{\mathfrak{H}_1}$ is self-adjoint \cite[p.9]{GG}. Therefore,
\begin{equation}\label{es121}
\mathfrak{H}=\mathfrak{H}_{0}\oplus\mathfrak{H}_1,
\end{equation}
where $\mathfrak{H}_{0}$ coincides with the closed linear span of all $\ker(S^*-\mu{I})$ ($\mu\in\mathbb{C}\setminus\mathbb{R}$) and the restriction $S_{0}:=S\upharpoonright_{\mathfrak{H}_{0}}$ is a simple symmetric operator in $\mathfrak{H}_{0}$.
It is clear that the restrictions $J\upharpoonright_{\mathfrak{H}_{0}}$ and $R\upharpoonright_{\mathfrak{H}_{0}}$ are
fundamental symmetries in $\mathfrak{H}_{0}$ and they satisfy
(\ref{es601}) for $S_0$. Applying  Theorem \ref{es600}, we establish the existence
of $J\upharpoonright_{\mathfrak{H}_{0}}$-self-adjoint extensions of $S_0$ with empty resolvent set.
Since an operator $A\in\Sigma_J$ has the decomposition $A=A_{0}\oplus{S_1}$ with respect to (\ref{es121}), where
$A_{0}$ is a $J\upharpoonright_{\mathfrak{H}_{0}}$-self-adjoint
extension of $S_{0}$, the set $\Sigma_J$  contains $J$-self-adjoint operators with empty resolvent set.
\end{remark}

From (\ref{es601}) one concludes that the four operators $I, J, R$, and $JR$ are linearly independent. Hence,
the operators $J$ and $R$ can be interpreted as  basis (generating) elements of the complex Clifford algebra
$$
{\mathcal C}l_2=\mbox{span}\{I, J, R, JR\}.
$$
\begin{corollary}\label{es105}
Let $S$ satisfy (\ref{es601})
and let $\widetilde{J}\in{\mathcal C}l_2$ be a nontrivial fundamental symmetry in $\mathfrak{H}$. Then there exists
$\widetilde{J}$-self-adjoint extensions of $S$ with empty resolvent set.
\end{corollary}
\emph{Proof.}
It is easy to see that an operator
$\widetilde{J}\in{\mathcal C}l_2$ is a nontrivial fundamental symmetry in $\mathfrak{H}$ (i.e., $\widetilde{J}^2=I$, $\widetilde{J}=\widetilde{J}^*$, and
$\widetilde{J}\not=I$) if and only if
\begin{equation}\label{es100}
\widetilde{J}=\alpha_{1}J+\alpha_{2}R+\alpha_{3}iJR, \quad  \alpha_1^2+\alpha_2^2+\alpha_3^2=1, \quad \alpha_j\in\mathbb{R}.
\end{equation}

Denote $\widetilde{R}=\beta_{1}J+\beta_{2}R+\beta_{3}iJR$, where $\sum\beta_j^2=1$, $\beta_j\in\mathbb{R}$.
By virtue of (\ref{es100}), $\widetilde{R}$ is a fundamental symmetry in $\mathfrak{H}$ which commutes with $S$.
Assuming $\sum\alpha_j\beta_j=0$, we obtain $\widetilde{J}\widetilde{R}=-\widetilde{R}\widetilde{J}$. Since $\widetilde{J}$ is a fundamental symmetry in $\mathfrak{H}$ which commutes with $S$, the statement follows from Theorem \ref{es600}. \rule{2mm}{2mm}
\subsubsection{The case $\textsf{Sh}\not\equiv{0}$.}
\begin{theorem}\label{es35}
Let $S$ be a simple symmetric operator with nonzero characteristic function
$\textsf{Sh}(\cdot)$ and the set $\Sigma_J$ contains operators with empty resolvent set. Then
all operators $C\in\mathfrak{U}$ have the form
  \begin{equation}\label{es602}
  C:={C}_{\chi,\omega}=J[(\cosh\chi)I+(\sinh\chi)R_\omega],
  \end{equation}
where $R$ satisfies (\ref{es601}), $R_{\omega}={R}e^{i{\omega}J}={R}[\cos\omega+i(\sin\omega)J]$, and  $\chi\in\mathbb{R}, \ \omega\in[0,2\pi)$.
\end{theorem}
\emph{Proof.}
First, we will show ${C}_{\chi,\omega}\in  \mathfrak{U}$.
Since $\Sigma_J$ contains operators with empty resolvent set, there exists a unitary mapping
$W : \mathfrak{H}_+\to\mathfrak{H}_-$ such that $S_+=W^{-1}S_-W$ (Corollary \ref{es202}). This allows one to determine
a fundamental symmetry $R$ in $\mathfrak{H}$  with the help of formula (\ref{es603}).

The operator $R$ possesses the properties (\ref{es601})
by the construction. Therefore, the subspaces $\mathfrak{N}_{\pm{i}}$ reduce $R$.
Let $\mathcal{R}_1=(r^1_{ij})_{i,j=1}^{2}$ and  $\mathcal{R}_2=(r_{ij}^2)_{i,j=1}^{2}$  be the matrix representations of $R\upharpoonright_{\mathfrak{N}_i}$ and $R\upharpoonright_{\mathfrak{N}_{-i}}$ with respect to the bases $e_{++}, e_{+-}$ and $e_{-+}, e_{--}$ of $\mathfrak{N}_{{i}}$  and $\mathfrak{N}_{{-i}}$, respectively. It follows from (\ref{es78}) and (\ref{es603}) that
$\mathcal{R}_j=\left(\begin{array}{cc} 0 & w_j^{-1} \\
w_j & 0 \end{array}\right)$, where $|w_1|=|w_2|=1$.   Moreover, since we assume (\ref{es498}), the parameter $\phi$ in the proof of
Corollary \ref{es202} is equal to zero and, hence, $w:=w_1=w_2$.
The exact values of unimodular constant $w$ depends on the choice of $W$.
Without loss of generality we may assume
(multiplying $W$ by an unimodular constant if it is necessarily) that $w=1$. Then
\begin{equation}\label{es700}
\mathcal{R}:=\mathcal{R}_1=\mathcal{R}_2=\left(\begin{array}{cc} 0 & 1 \\
1 & 0 \end{array}\right).
\end{equation}

Let us consider the collections of operators $C_{\chi,\omega}$ determined by (\ref{es602})
It is known that $C_{\chi,\omega}=Je^{\chi{R_\omega}}$, where $R_{\omega}={R}e^{i{\omega}J}={R}[\cos\omega+i(\sin\omega)J]$
is a fundamental symmetry in $\mathfrak{H}$, which anticommutes with $J$  (i.e., $R_{\omega}J=-JR_{\omega}$)  \cite{AKG}.
Such a representation leads to the conclusion that $C_{\chi,\omega}^2=I$ and $JC_{\chi,\omega}>0$. Moreover
$SC_{\chi,\omega}=SC_{\chi,\omega}$ due to (\ref{es12}) and (\ref{es601}).
Therefore, an arbitrary $C_{\chi,\omega}$ belongs to $\mathfrak{U}$.

Rewriting (\ref{es602}) as follows
$$
{C}_{\chi,\omega}=(\cosh\chi){J}+(\sinh\chi)(\cos\omega){JR}-i(\sinh\chi)(\sin\omega){R}
$$
and using (\ref{es700}) we obtain that both matrix
representations of $C_{\chi,\omega}\upharpoonright_{\mathfrak{N}_i}$ and of $C_{\chi,\omega}\upharpoonright_{\mathfrak{N}_{-i}}$ coincide
with
$$
\mathcal{C}_{\chi,\omega}=\left(\begin{array}{cc}
\cosh\chi & (\sinh\chi)e^{-i\omega} \\
-(\sinh\chi)e^{i\omega} & -\cosh\chi
\end{array}\right).
$$

Let $C\in\mathfrak{U}$. Then the matrix representations of its restrictions
$C\upharpoonright_{\mathfrak{N}_{{i}}}$ and $C\upharpoonright_{\mathfrak{N}_{`�i}}$ coincide with $\mathcal{C}_{\chi_1,\omega_1}$ and $\mathcal{C}_{\chi_2,\omega_2}$ defined by (\ref{es124}). Furthermore, since $\textsf{Sh}(\mu)C=C\textsf{Sh}(\mu)$ (see (\ref{es407})), the identity  (\ref{es408}) holds. That is equivalent to the relations
$\chi_1=\chi_2$ and $e^{-i\omega_1}=e^{-i\omega_2}$ (since (\ref{es498}) is true and $s_+\not\equiv{0}$).

Setting $\chi=\chi_1=\chi_2$ and $\omega=\omega_1$, one concludes
that the matrix representations $\mathcal{C}_{\chi_j, \omega_j}$  coincides with
$\mathcal{C}_{\chi,\omega}$. Therefore, $C=C_{\chi,\omega}$ due to
Lemma \ref{neww41}. Thus, the collection of operators $\{C_{\chi,\omega}\}$ defined by (\ref{es602}) coincides with $\mathfrak{U}$.
 \rule{2mm}{2mm}

Combining Theorem \ref{es35} with \cite[Theorem 3.2, Proposition 3.3]{AKG}, we immediately derive the following
statement.

\begin{corollary}\label{es303}
Let $S$ and $\Sigma_J$ satisfy the condition of Theorem \ref{es35}
and let $A_{U}\in\Sigma_J$ be defined by (\ref{eee6}) - (\ref{aaa5}). Then
the strict inclusions
$$\Upsilon_{\mathfrak{U}}\subset\Upsilon_{J}\subset\Sigma_{J}^{st}$$
 hold and the following
relations are true.
\begin{enumerate}
  \item[(i)] $A_{U}$ belongs to $\Upsilon_{\mathfrak{U}}$ if and only if
$$
U=e^{i\frac{\pi}{2}}\left(\begin{array}{cc} 0 & e^{i\xi}  \\
-e^{-i\xi} & 0 \end{array}\right), \quad \xi\in{[0,2\pi)};$$
 \item[(ii)] $A_{U}$ belongs to $\Upsilon_J$ if and only if
$$
U=e^{i\phi}\left(\begin{array}{cc} 0 & e^{i\xi} \\
-e^{-i\xi} & 0 \end{array}\right), \quad \phi, \xi\in{[0,2\pi)};
$$
 \item[(iii)] $A_{U}$ belongs to $\Sigma_J^{st}\setminus\Upsilon_J$ if and only if
$$
U=e^{i\phi}\left(\begin{array}{cc} qe^{i\gamma} & re^{i\xi} \\
-re^{-i\xi} & qe^{-i\gamma} \end{array}\right), \quad  \gamma, \xi\in{[0,2\pi)},\ q,r > 0,\ q^2+r^2=1,
$$
\end{enumerate}
where $0<q<|\cos\phi|$. In that case the operator $A_{U}$ has the $C_{\chi,\omega}$-symmetry,
where $\omega=\gamma$ and $\chi$ is determined by the relation $q=-\tanh\chi\cos\phi.$
\end{corollary}

\subsubsection{The case  $\textsf{Sh}\equiv{0}$.} $\mbox{}$  \\

If $\textsf{Sh}\equiv{0}$, then ${s}_+(\mu)={s}_-(\mu)=0$ for all $\mu\in\mathbb{C}_+$.
Therefore, by Theorem \ref{es95}, $\Sigma_J$ contains operators with empty resolvent set and Theorem \ref{es600} and Corollary
\ref{es105} hold. However Theorem \ref{es35} is not true due to the fact that the set of all
stable $C$-symmetries $\mathfrak{U}$ is much more greater then the formula (\ref{es602}) provides.
That is why the commutation condition (\ref{es408}) is vanished for ${s}_\pm \equiv{0}$ and we cannot establish
the relationship between parameters $\chi_1, \omega_1$ and $\chi_2, \omega_2$ of matrices $\mathcal{C}_{\chi_j,\omega_j}$
(see the proof of Theorem \ref{es35}).
\begin{theorem}\label{es2020}
Let $S$ be a simple symmetric operator with zero characteristic function and
let $A_{U}\in\Sigma_J$ be defined by (\ref{eee6}) - (\ref{aaa5}).
Then $\Upsilon_{\mathfrak{U}}=\emptyset$ and
the strict inclusions
$$\Upsilon_{\mathfrak{U}}\subset\Upsilon_{J}\subset\Sigma_{J}^{st}$$
 hold.
\begin{enumerate}
  \item[(i)] $A_U$ belongs to $\Upsilon_J$ if and only if
$$
U=e^{i\phi}\left(\begin{array}{cc} 0 & e^{i\xi} \\
-e^{-i\xi} & 0 \end{array}\right), \quad \phi, \xi\in{[0,2\pi)};
$$
  \item[(ii)] $A_U$ belongs to $\Sigma_J^{st}\setminus\Upsilon_J$ if and only if
$$
U=e^{i\phi}\left(\begin{array}{cc} qe^{i\gamma} & re^{i\xi} \\
-re^{-i\xi} & qe^{-i\gamma} \end{array}\right), \quad  \phi,\gamma,{\xi}\in{[0,2\pi)},\  q,r > 0,\  q^2+r^2=1.
$$
\end{enumerate}
\end{theorem}
\emph{Proof.} (i) It follows from \cite[Proposition 3.3]{AKG}.

(ii) Let $A_U\in\Sigma_{J}^{st}$. Then  $A_{U}C=CA_{U}$ for  some choice of $C\in\mathfrak{U}$. This is equivalent to the relation
$CM=M$, where $M=\mbox{span}\{d_1,d_2\}$ is defined by (\ref{e6}) and (\ref{aaa5}) (see the proof of Theorem \ref{es112}).
Moreover, it follows from the proof of Theorem \ref{es112} that the operators
$C\upharpoonright_{\mathfrak{N}_{{i}}}$  and $C\upharpoonright_{\mathfrak{N}_{{-i}}}$ acts in ${\mathfrak{N}_{{i}}}$
and ${\mathfrak{N}_{{-i}}}$, respectively and they have the matrix representations
$\mathcal{C}_{\chi_1,\omega_1}$ and $\mathcal{C}_{\chi_2,\omega_2}$ defined by formula (\ref{es124}).

Combining \cite[Lemma 3.3]{KU} with Lemma \ref{neww41} we conclude that
the correspondence
\begin{equation}\label{nenede1}
{{C}\in\mathfrak{U}}\rightarrow\{\mathcal{C}_{\chi_1,\omega_1}, \mathcal{C}_{\chi_2,\omega_2}\}, \qquad \chi_j\in\mathbb{R}, \ \omega_j\in[0,2\pi)
\end{equation}
is \emph{bijective}
for the case of zero characteristic function ($\textsf{Sh}\equiv{0}$).

It follows from (\ref{aaa5}) and (\ref{es124}) that
\begin{eqnarray*}
Cd_1=\mathcal{C}_{\chi_1,\omega_1}e_{++}+qe^{i(\phi+\gamma)}\mathcal{C}_ {\chi_1,\omega_1}e_{+-}+re^{i(\phi+\xi)}\mathcal{C}_{\chi_2,\omega_2}e_{-+}= &  & \\
k_1e_{++}-[\sinh\chi_1e^{i\omega_1}+qe^{i(\phi+\gamma)}\cosh\chi_1]e_{+-}+[re^{i(\phi+\xi)}\cosh\chi_2]e_{-+}+k_2e_{--}, & &
\end{eqnarray*}
where
\begin{equation}\label{nenede3}
k_1=\cosh\chi_1+qe^{i(\phi+\gamma)}\sinh\chi_1e^{-i\omega_1}, \ k_2=-re^{i(\phi+\xi)}\sinh\chi_2e^{i\omega_2}.
\end{equation}

Taking the definition (\ref{aaa5}) of $d_j$ into account we conclude that $Cd_1\in{M}$ if and only if
$Cd_1=k_1d_1+k_2d_2$, where $k_j$ are defined by (\ref{nenede3}). A direct calculation shows that the last identity holds if we set
\begin{equation}\label{nenede2}
\chi=\chi_1=\chi_2=-\tanh^{-1}q, \quad \omega_1=\frac{\gamma+\phi}{2}, \quad \omega_2=\frac{\gamma-\phi}{2}.
\end{equation}

A similar reasoning shows that $Cd_2\in{M}$ if we choose parameters $\chi_j$ and $\omega_j$ according to (\ref{nenede2}).
Note that $\chi$ can be defined in (\ref{nenede2}) just in the case $0\leq{q}<1$.

Thus, if $A_{U}\in\Sigma_J$ is defined by (\ref{eee6}) - (\ref{aaa5}) with $0\leq{q}<1$, then choosing parameters $\chi_j$, $\omega_j$
due to (\ref{nenede2}) and using the bijection (\ref{nenede1}), we establish the existence of $C\in\mathcal{U}$ such that $A_UC=CA_U$.
Therefore $A_U\in\Sigma_J^{st}$. Since $A_U\in\Upsilon_J$  when $q=0$ (see item (i)) and the spectrum of $A_U$ coincides with
$\mathbb{C}$ when $q=1$ (it follows from Corollary \ref{es200}), we prove (ii).

Let us assume  that $A_U\in\Upsilon_{\mathfrak{U}}$. In that case
$A_{U}C=CA_{U}$ for all $C\in\mathfrak{U}$. Taking (\ref{nenede1}) into account, we conclude that the element
$Cd_1=\mathcal{C}_{\chi_1,\omega_1}e_{++}+qe^{i(\phi+\gamma)}\mathcal{C}_ {\chi_1,\omega_1}e_{+-}+re^{i(\phi+\xi)}\mathcal{C}_{\chi_2,\omega_2}e_{-+}$ belongs to $M$ (i.e., $Cd_1=k_1d_1+k_2d_2$, where $k_j$ are defined by (\ref{nenede3})) for \emph{all values} of parameters $\chi_j$ and $\omega_j$. This is impossible.
Hence, $\Upsilon_{\mathfrak{U}}=\emptyset$.
\rule{2mm}{2mm}

The next statement is a direct consequence of Proposition \ref{sese1} and Theorem \ref{es2020}.
\begin{corollary}[ \cite{KU}]
If $S$ is a simple symmetric operator with zero characteristic function, then an operator
$A_U\in\Sigma_J$ has real spectrum if and only if $A_U$ has stable $C$-symmetry and, hence,
$A_U$ is similar to self-adjoint operator. Otherwise, the spectrum of $A_U$ coincides with
$\mathbb{C}$.
\end{corollary}

\section{Examples}
\subsection{Degenerate Sturm-Liouville problems on the finite interval.}
The necessary and sufficient conditions for the Dirichlet eigenvalue problem associated with
the Sturm-Liouville equation
\begin{equation}\label{dedede4}
-(p(x)y')'=\lambda{r}(x)y, \qquad -\infty<a\leq{x}\leq{b}<\infty
\end{equation}
to be degenerate (i.e., the spectrum of this eigenvalue problem fills the whole complex plane) were established in \cite{MIN}. We consider one of the simplest cases where $p(x)=r(x)=(\textsf{sgn}\ x)$ and $[a,b]=[-1,1]$.

The symmetric operator $S$ associated with $-(\textsf{sgn}\ x)((\textsf{sgn}\ x)y')'$ and boundary conditions
$y(-1)=y(1)=0$ takes the form $Sy=-y''$,
\begin{equation}\label{did345}
\mathcal{D}(S)=\{y\in{W}^2_2(-1,0)\oplus{W}^2_2(0,1) \ | \ y(0\pm)=y'(0\pm)=y(\pm1)=0\}
\end{equation}
and (\ref{dedede4}) takes the form $Sy=\lambda{y}$.

The operator $S$ has the deficiency indices $<2,2>$ and it commutes with the fundamental symmetry $Jy(x)=(\textsf{sgn}\ x)y(x)$ in $\mathfrak{H}=L_2(-1,1)$.
The corresponding symmetric operators $S_\pm{y}=-y''$ (see (\ref{ea1b})) with the domains
$$
\begin{array}{c}
\mathcal{D}(S_+)=\{y\in{W}^2_2(0,1) \ | \ y(0+)=y'(0+)=y(1)=0\}, \vspace{3mm} \\
\mathcal{D}(S_-)=\{y\in{W}^2_2(-1,0) \ | \ y(0-)=y'(0-)=y(-1)=0\}
\end{array}
$$
act in $\mathfrak{H}_+=L_2(0,1)$ and $\mathfrak{H}_-=L_2(-1,0)$, respectively.

Consider the parity operator $\mathcal{P}y(x)=y(-x)$ and set $R:=\mathcal{P}$. It is clear that $R$ is a fundamental symmetry in $L_2(-1,1)$ and it satisfies (\ref{es601}).
To describe these operators we observe that solutions
$y_\mu^{\pm}(x)$ of the equations
$$
S^*_\pm{y}-\mu{y}=-y''(x)-\mu{y(x)}=0, \quad y(\pm{1})=0, \quad \mu\in\mathbb{C}_+
$$
have the form
$$
y_\mu^{+}(x)=\left\{\begin{array}{ll}
\sin\sqrt\mu(x-1), &  x\in[0,1] \\
0,  &  x\in[-1,0],
\end{array}\right.
$$
$$
y_\mu^{-}(x)=\left\{\begin{array}{ll}
0,  & x\in[0,1] \\
-\sin\sqrt\mu(x+1) & x\in[-1,0].
\end{array}\right.
$$
Here $\sqrt{\cdot}$ denotes the
branch of the square root defined in
$\mathbb C$ with a cut along $[0,\infty)$ and fixed by Im$\,\sqrt{\lambda}>0$ if $\lambda\not\in [0,\infty)$. Moreover, $\sqrt{\cdot}$
is continued to $[0,\infty)$ via $\lambda \mapsto \sqrt{\lambda} \geq 0$ for
$\lambda\in [0,\infty)$.
According to (\ref{ura1}), the elements $e_{\pm\pm}$ can be chosen as follows:
$$
e_{++}=y_i^{+}, \quad e_{+-}=y_i^{-}, \quad e_{-+}=y_{-i}^{+}, \quad e_{--}=y_{-i}^{-}
$$
and the functions $s_\pm(\mu)$
in (\ref{GDR1}) can be calculated immediately by repeating the arguments in \cite{SH}.
For completeness we outline the method.

The characteristic function $\textsf{Sh}_+(\mu)$ of $S_+$ is determined by
the first relations in (\ref{GDR1}) and (\ref{GDR3}). Employing here
(\ref{GDR2}) we get
\begin{equation}\label{ura2}
y_\mu^{+}(x)=u(x)+ce_{++}-cs_+(\mu)e_{-+}, \quad u\in\mathcal{D}(S_+), \quad x\in[0,1],
\end{equation}
where $c$ is a constant which is easily determined by setting $x=0$ and taking into account
the relevant boundary conditions:
$$
c=\frac{\sin\sqrt{\mu}}{\sin\sqrt{i}-s_+(\mu)\sin\sqrt{-i}}.
$$
Differentiating (\ref{ura2}) with a subsequent setting $x=0$ gives rise to
$$
\sqrt{\mu}\cos\sqrt{\mu}=c\sqrt{i}\cos\sqrt{i}-cs_+(\mu)\sqrt{-i}\cos\sqrt{-i}.
$$
The last two relations leads to the conclusion:
$$
s_+(\mu)=\frac{\sqrt{i}\sin\sqrt{\mu}\cos\sqrt{i}-\sqrt{\mu}\cos\sqrt{\mu}\sin\sqrt{i}}{\sqrt{-i}\sin\sqrt{\mu}\cos\sqrt{-i}-\sqrt{\mu}\cos\sqrt{\mu}\sin\sqrt{-i}}.
$$

Considering the characteristic function $\textsf{Sh}_-$ of $S_-$ we obtain the same expression
for $s_-(\mu)$. Thus $s_+=s_-\not\equiv{0}$. By Theorem \ref{es95},
the set $\Sigma_J$ of $J$-self-adjoint extensions of $S$ contains operators with empty resolvent set. Applying Corollary \ref{es200} and taking the explicit form of elements $e_{\pm\pm}$ into account we derive the following description of all possible $J$-self-adjoint extensions $A(=A_\gamma)$ of $S$ with empty resolvent set:
$A_\gamma{y}=-y'',$
$$
\mathcal{D}(A_\gamma)=\left\{\,y\in{W}^2_2(-1,0)\oplus{W}^2_2(0,1) \ \left|\right. \ \begin{array}{l}
e^{i\gamma}y(0+)=y(0-), \vspace{2mm} \\
e^{i\gamma}y'(0+)=-y'(0-),\vspace{2mm} \\
y(\pm1)=0,
\end{array}\right\}
$$
where $\gamma\in[0,2\pi)$ is an arbitrary parameter.

\begin{remark}
Since the symmetric operator $S$ satisfies (\ref{es601}) with fundamental symmetries $J=(\textsf{sgn}\ x)I$ and $R=\mathcal{P}$, Corollary \ref{es105} implies the existence of $\widetilde{J}$-self-adjoint extensions of $S$
with empty resolvent set for any nontrivial fundamental symmetry $\widetilde{J}$
which belongs to the Clifford algebra ${\mathcal C}l_2=\mbox{span}\{I, J, R, JR\}$.
\end{remark}

\subsection{Indefinite Sturm-Liouville operators $(\textsf{sgn}\ x)(-\frac{d^2}{dx^2} +q(x))$.}
Consider the indefinite Sturm-Liouville differential expression
$$
a(y)(x)=(\textsf{sgn}\ x)(-y''(x)+q(x)y(x)), \quad x\in\mathbb{R}
$$
with a real potential $q\in{L^1_{loc}(\mathbb{R})}$ and denote by $\mathfrak{D}$ the set of all functions $y\in{L_2(\mathbb{R})}$ such that $y$ and $y'$ are absolutely continuous and $a(y)\in{L_2(\mathbb{R})}$.
On $\mathfrak{D}$ we define the operator  $A$ as follows:
\begin{equation}\label{es500}
 Ay=a(y), \quad \mathcal{D}(A)=\mathfrak{D}.
\end{equation}

Assume in what follows the limit point case of $a(y)$ at both $-\infty$ and $+\infty$. Then we obtain
the $J$-self-adjoint operator $A$ in the Krein space $(L_2(\mathbb{R}), [\cdot,\cdot]_J)$,
where $J=(\textsf{sgn}\ x)I$.

The operator $A$ is a $J$-self-adjoint extension of the symmetric operator
\begin{equation}\label{es787}
S=(\textsf{sgn}\ x)\left(-\frac{d^2}{dx^2}+q\right), \quad \mathcal{D}(S)=\{y\in\mathfrak{D} \  | \ y(0)=y'(0)=0\}
\end{equation}
with deficiency indices $<2,2>$.

The operator $S$ commutes with $J$ and its restrictions onto subspaces $L_2(\mathbb{R}_\pm)$ of the
fundamental decomposition $L_2(\mathbb{R})=L_2(\mathbb{R}_+)\oplus{L_2(\mathbb{R}_-)}$ coincides with
the symmetric operators
$$
S_+=-\frac{d^2}{dx^2}+q_+, \quad S_-=\frac{d^2}{dx^2}-q_-, \quad \mathcal{D}(S_\pm)=P_{\pm}\mathcal{D}(S), \quad
q_\pm=q\upharpoonright_{\mathbb{R}_\pm}
$$
with deficiency indices $<1,1>$ acting in the Hilbert spaces $\mathfrak{H}_+=L_2(\mathbb{R}_+)$ and $\mathfrak{H}_-=L_2(\mathbb{R}_-)$, respectively and $P_\pm$ are orthogonal projectors onto $L_2(\mathbb{R}_\pm)$ in
$L_2(\mathbb{R})$.

Denote by $c_\mu(\cdot)$, $s_\mu(\cdot)$  the solutions of the equation
 $$
 -f''(x)+q(x)f(x)={\mu}f(x), \qquad  x\in\mathbb{R}, \quad \mu\in\mathbb{C}
 $$
 with the boundary conditions
 \begin{equation}\label{es654}
 c_\mu(0)=s_\mu'(0)=1, \qquad c_\mu'(0)=s_\mu(0)=0.
 \end{equation}

Due to the limit point case at $\pm\infty$ there exist unique holomorphic functions
$M_\pm(\mu)$ ($\mu\in\mathbb{C}\setminus\mathbb{R}$) such that
the functions
\begin{equation}\label{es709}
\psi_\mu^{\pm}(x)=\left\{\begin{array}{lr}
s_{\pm\mu}(x)-M_{\pm}(\mu)c_{\pm\mu}(x), &  x\in\mathbb{R}_{\pm} \\
0, & x\in\mathbb{R}_{\mp}
\end{array}\right.
\end{equation}
belongs to $L_2(\mathbb{R})$. The functions  $M_\pm(\cdot)$  are called the Titchmarsh-Weyl coefficients of the differential expression $a(\cdot)$ on $\mathbb{R}_\pm$ (see, e.g., \cite[Definition 2.2]{KT}). They are Nevanlinna functions and they
satisfy the following asymptotic behavior
\begin{equation}\label{es534}
M_\pm(\mu)=\pm\frac{i}{\sqrt{\pm{\mu}}}+O\left(\frac{1}{|\mu|}\right),  \ (\mu\to\infty, 0<\delta<\mbox{arg}\ \mu<\pi-\delta)
\end{equation}
for $\delta\in(0,\frac{\pi}{2})$, see \cite{EV}.

The asymptotic behavior (\ref{es534}) was used for justifying the property $\rho(A)\not=\emptyset$
for the concrete  $J$-self-adjoint extension $A$ of $S$ defined by (\ref{es500}), cf.\ \cite{KarMal}.
We extend this result for all operators in $\Sigma_J$.
\begin{theorem}\label{es902}
Let the symmetric operator $S$ be defined by (\ref{es787}) and $J=(\textsf{sgn}\ x)I$.
Then the set $\Sigma_J$ of $J$-self-adjoint extensions of $S$  does not contain
operators with empty resolvent set.
\end{theorem}
\emph{Proof.} The proof is divided into two steps. In the first one we calculate the characteristic function
of $S$. In step 2 we apply Theorem \ref{es95}.

{\bf Step 1.} It follows from the definition of $S_\pm$ and (\ref{es709}) that
the defect subspaces $\mathfrak{N}_{\pm{i}}(S_+)$ coincides with $\mbox{span}\{\psi_{\pm{i}}^{+}\}$ and
the defect subspaces $\mathfrak{N}_{\pm{i}}(S_-)$ coincides with $\mbox{span}\{\psi_{\pm{i}}^{-}\}$.
Therefore, we can choose basis elements $\{e_{\pm\pm}\}$ as follows:
$$
e_{++}=\psi_{i}^{+}, \quad e_{-+}=\psi_{-i}^{+}, \quad e_{+-}=c\psi_{{i}}^{-}, \quad e_{--}=c\psi_{-i}^{-},
$$
where an auxiliary constant $c>0$ is determined by the condition $\|\psi_{i}^{+}\|=\|\psi_{i}^{-}\|$
(or, that is equivalent, by the condition $\|\psi_{-i}^{+}\|=\|\psi_{-i}^{-}\|$)
and it ensures the equality of norms $\|e_{++}\|=\|e_{+-}\|=\|e_{-+}\|=\|e_{--}\|$.

By virtue of (\ref{es654}) and (\ref{es709}),
\begin{equation}\label{dedede5}
e_{\pm+}(0)=-M_{+}(\pm{i}), \ e_{\pm+}'(0)=1,  \ e_{\pm-}(0)=-cM_{-}(\pm{i}), \ e_{\pm-}'(0)=c.
\end{equation}
Using these boundary conditions and repeating the arguments of Subsection 5.1, we arrive at the conclusion that
the characteristic function $\textsf{Sh}$ of $S$ is defined by the following functions $s_+(\cdot)$ and $s_-(\cdot)$ in (\ref{GDR1}):
\begin{equation}\label{es480}
s_+(\mu)=\frac{M_+(\mu)-M_+(i)}{M_+(\mu)-M_+(-i)}, \qquad s_-(\mu)=\frac{M_-(\mu)-M_-(i)}{M_-(\mu)-M_-(-i)}.
\end{equation}

{\bf Step 2.} By Theorem \ref{es95} the set $\Sigma_J$ contains operators with empty resolvent set if and only if
$e^{2i\phi}s_-(\mu)=s_+(\mu)$, $\mu \in \mathbb C_+$, for a certain choice of $\phi\in[0,2\pi)$.
Tending $\mu\to\infty$ in this identity and taking (\ref{es534}) and (\ref{es480}) into account, we obtain that
\begin{equation}\label{es45}
e^{2i\phi}=\frac{M_+(i)M_-(-i)}{M_+(-i)M_-(i)}.
\end{equation}
Rewriting $e^{2i\phi}s_-(\mu)=s_+(\mu)$ with the use of (\ref{es480}) and (\ref{es45}) we get
\begin{eqnarray}\label{es46}
M_+(\mu)M_-(\mu)[e^{2i\phi}-1]+M_+(\mu)[M_-(-i)-e^{2i\phi}M_-(i)] & & \nonumber \\ +M_-(\mu)[M_+(i)-e^{2i\phi}M_+(-i)]=0. & &
\end{eqnarray}

Denote $M_{\pm}(i)=e^{i\theta_{\pm}}|M_{\pm}(i)|$, where $\theta_\pm\in(0,\pi)$ (since $\mbox{Im} \ M_\pm(i)>0$). Then (\ref{es45}) takes the form $e^{2i\phi}=e^{2i(\theta_+-\theta_-)}$ and relation (\ref{es46}) can
rewriting (after routine transformations) as follows:
\begin{eqnarray}\label{es46b}
M_+(\mu)M_-(\mu)\sin(\theta_+-\theta_-)-M_+(\mu)|M_-(i)|\sin\theta_+ & &  \\
+M_-(\mu)|M_+(i)|\sin\theta_-=0. & \forall{\mu}\in{C}_+ & \nonumber
\end{eqnarray}
Since the coefficients $|M_\mp(i)|\sin\theta_\pm$ of $M_\pm(\mu)$ are real, identity (\ref{es46b})
cannot be true for the whole $\mathbb{C}_+$ (due to the asymptotic behavior (\ref{es534})).
Therefore, $\Sigma_J$ does not contain operators with empty resolvent set.
\rule{2mm}{2mm}
\begin{remark} The relation (\ref{es46b}) is reduced to $M_+\equiv{M}_-$, if
we are trying to prove $\rho(A)\not=\emptyset$ for the concrete operator $A$ defined by (\ref{es500}), see, e.g., \cite[Proposition 2.5]{KarMal}.
This could also be deduced here by a simple calculation involving
 Corollary \ref{es200} and  (\ref{es400}).
\end{remark}
By virtue of Theorems \ref{es112}, \ref{es902}
the set $\Sigma_J^{st}$ of $J$-self-adjoint operators with stable $C$-symmetry
is reduced to the set $\Upsilon_J$ of self-adjoint extensions of $S$ which
commute with $J$ in the case of indefinite Sturm-Liouville operators.
The set $\Upsilon_J$ consists of all self-adjoint
extensions of $S$ with \emph{separated boundary conditions on $0$}, i.e.,
$$
A\in\Upsilon_J \iff Ay=a(y), \quad \mathcal{D}(A)=\{y\in\mathfrak{D} \ | \ {a}_{\pm}f(0\pm)-{b}_{\pm}f'(0\pm)=0\}.
$$


\subsection{One dimensional impulse operator with point perturbation.}
Consider the symmetric operator
$$
S=-i\frac{d}{dx}, \quad \mathcal{D}(S)=\{y\in{W}^1_2(\mathbb{R}, \mathbb{C}^2) \ | \ y(0)=0\}
$$
in the Hilbert space $L_2(\mathbb{R}, \mathbb{C}^2):=L_2(\mathbb{R})\otimes{\mathbb{C}^2}$.
\begin{lemma}\label{es490}
The operator $S$ has deficiency indices $<2,2>$ and its characteristic function $\textsf{Sh}$
is equal to zero.
\end{lemma}
\emph{Proof.}
The operator $S$ can be presented $S=S_1+S_2$ with respect to the decomposition
$L_2(\mathbb{R}, \mathbb{C}^2)=L_2(\mathbb{R}_-, \mathbb{C}^2)\oplus{L_2(\mathbb{R}_+, \mathbb{C}^2)}$.
The restrictions $S_1=S\upharpoonright_{L_2(\mathbb{R}_-, \mathbb{C}^2)}$ and
$S_2=S\upharpoonright_{L_2(\mathbb{R}_+, \mathbb{C}^2)}$ are maximal symmetric operators in the Hilbert
spaces $L_2(\mathbb{R}_-, \mathbb{C}^2)$ and $L_2(\mathbb{R}_+, \mathbb{C}^2)$, respectively, with deficiency indices $<0,2>$ and $<2,0>$, respectively.
Therefore $S$ has deficiency indices $<2,2>$ and $\mathfrak{N}_{\mu}(S)=\mathfrak{N}_{\mu}(S_2)$ for all $\mu\in\mathbb{C}_+$ (since $S_2$ has
deficiency indices $<2,0>$). An arbitrary $f_\mu\in\mathfrak{N}_{\mu}(S)$ admits the representation
$$
f_\mu=u+f_i, \quad  u\in\mathcal{D}(S_2), \quad f_i\in\mathfrak{N}_{i}(S_2).
$$
Comparing the obtained formula with (\ref{GDR2}) we obtain $\textsf{Sh}(\mu)=0$.
\rule{2mm}{2mm}

\begin{remark}
The operator $J=(\textsf{sgn}\ x)I$ is a fundamental symmetry in $L_2(\mathbb{R}, \mathbb{C}^2)$
and $S$ commutes with $J$. The symmetric operators $S_-$ and $S_+$  in (\ref{ea1b}) coincides with
$S_1$ and $S_2$, respectively, and hence their deficiency indices are $<0,2>$ and $<2,0>$. Hence,  there are no $J$-self-adjoint extensions of $S$ and
 the sets $\Sigma_J$ and $\Upsilon_J$  are empty.
\end{remark}

To achieve a non-empty set $\Sigma_J$, we have to choose a fundamental symmetry $J$ in such a way
that the deficiency indices of $S_\pm$ in (\ref{ea1b}) are $<1,1>$. To this end, we
write an arbitrary element $y\in{L_2(\mathbb{R}, \mathbb{C}^2)}$ as follows
$$
y=\left(\begin{array}{c}
y_1 \\
y_2 \end{array}
\right)=y_1\otimes{h_+}+y_2\otimes{h_-}, \quad h_+=\left(\begin{array}{c}
1 \\
0 \end{array}
\right), \  h_-=\left(\begin{array}{c}
0 \\
1\end{array}\right)
$$
and consider the fundamental symmetry
$Jy=\left(\begin{array}{c}
y_1 \\
-y_2 \end{array}
\right)$ in $L_2(\mathbb{R}, \mathbb{C}^2)$.
In that case, the operators $S_\pm$ in (\ref{ea1b}) act in the Hilbert spaces
$L_2(\mathbb{R}, \mathcal{H}_\pm)$, where $\mathcal{H}_\pm=\mbox{span}\{h_\pm\}$ and they are determined by the formulas
\begin{equation}\label{es590}
S_\pm=-i\frac{d}{dx}, \quad \mathcal{D}(S_\pm)=\{y\in{W}^1_2(\mathbb{R}, \mathcal{H}_\pm) \ | \ y(0)=0 \}.
\end{equation}
Obviously, $S_\pm$ have deficiency indices $<1,1>$. This means that the set $\Sigma_J$ is non-empty and its
elements can be parameterized by unitary matrices $U$ in (\ref{eee6}).

In order to describe the subset of $J$-self-adjoint extensions with empty resolvent set in
$\Sigma_J$ we have to calculate basis elements $\{e_{\pm\pm}\}$ (see (\ref{ura1})) and to apply
Corollary \ref{es200}.

Denote by
$$
y_{i}(x)=\left\{\begin{array}{lr}
e^{-x}, &  x\geq{0} \\
0, & x<0,
\end{array}\right. \qquad  y_{-i}(x)=\left\{\begin{array}{lr}
0, &  x\geq{0} \\
e^x, & x<0
\end{array}\right.
$$
the solutions of the equation $-iy'-\mu{y}=0$ \ ($\mu\in\{i,-i\}$).
Using the definition of $S_\pm$ and (\ref{ura1}) we obtain
$$
e_{++}=y_i\otimes{h_+}, \quad e_{+-}=y_i\otimes{h_-}, \quad e_{-+}=y_{-i}\otimes{h_+}, \quad e_{--}=y_{-i}\otimes{h_-}.
$$

Corollary \ref{es200} and equalities (\ref{e6}), (\ref{aaa5}) imply that an
arbitrary $J$-self-adjoint extension $A_U$ with empty resolvent set has the domain $\mathcal{D}(A_U)=\mathcal{D}(S)\dot{+}M$, where $M$ is a linear span of elements
$$
d_1=e_{++}+e^{i(\phi+\gamma)}e_{+-}, \quad d_2=e_{--}+e^{i(\phi-\gamma)}e_{-+}, \quad \phi,\gamma\in[0,2\pi).
$$
The obtained expression leads (after some trivial calculations) to the following description of $J$-self-adjoint extensions $A_U(=A_{\phi\gamma})$ of $S$ with empty resolvent set:
$A_{\phi\gamma}{y}=-iy'$,
$$
\mathcal{D}(A_{\phi\gamma})=\left\{y=\left(\begin{array}{c}
y_1  \\
y_2
\end{array}\right)\in{W}_2^1(\mathbb{R}\setminus\{0\})\otimes{\mathbb{C}}^2 \
 \left|\right. \ \begin{array}{l}
y_2(0+)=e^{i(\gamma+\phi)}y_1(0+) \vspace{2mm} \\
y_2(0-)=e^{i(\gamma-\phi)}y_1(0-)
\end{array}\right\},
$$
where $\phi,\gamma\in[0,2\pi)$ are arbitrary parameters.

\subsection{One dimensional Dirac operator with point perturbation.}
Let us consider the free Dirac operator $D$ in the space
$L_2(\mathbb{R})\otimes{\mathbb{C}}^2$:
$$
D=-ic\frac{d}{dx}\otimes{\sigma_1}+\frac{c^2}{2}\otimes\sigma_3,
\quad \mathcal{D}(D)={W}_2^1(\mathbb{R})\otimes{\mathbb{C}}^2,
$$
where $\sigma_1=\left(\begin{array}{cc} 0  & 1 \\
1 & 0
\end{array}\right)$, \ $\sigma_3=\left(\begin{array}{cc} 1  & 0 \\
0 & -1
\end{array}\right)$ are Pauli matrices and $c>0$.

The symmetric Dirac operator
$$
S=D\upharpoonright{\{u\in{W}_2^1(\mathbb{R})\otimes{\mathbb{C}}^2
\ | \ u(0)=0\}}
$$
 has the deficiency indices $<2,2>$, see \cite{AL}, and
it commutes with the fundamental symmetry
$J={\mathcal{P}}\otimes{\sigma_3}$ in
$L_2(\mathbb{R})\otimes{\mathbb{C}}^2$, where $\mathcal{P}$ is
the parity operator $\mathcal{P}y(x)=y(-x)$.
In that case, the operators $S_\pm$ in (\ref{ea1b}) are restrictions of $S$
onto the Hilbert spaces
$$
[L_2^{even}(\mathbb{R})\otimes\mathcal{H}_+]\oplus[{L_2^{odd}(\mathbb{R})\otimes\mathcal{H}_-}], \quad [L_2^{odd}(\mathbb{R})\otimes\mathcal{H}_+]\oplus[{L_2^{even}(\mathbb{R})\otimes\mathcal{H}_-}],
$$
respectively ($\mathcal{H}_\pm$ are the same as in (\ref{es590})) and $S_\pm$ have deficiency indices $<1, 1>$.

The defect subspaces $\mathfrak{N}_{i}$ and $\mathfrak{N}_{-i}$ of
$S$ coincide, respectively, with the linear spans of
the functions $\{y_{1+}, y_{2+}\}$ and $\{y_{1-}, y_{2-}\}$, where
\begin{equation}\label{aaa14}
y_{1\pm}(x)=\left(\begin{array}{c}
  -ie^{{\mp{i}t}} \\
 (\textsf{sgn}\ x)
 \end{array}\right){e^{i\tau|x|}}, \quad
 y_{2\pm}(x)=(\textsf{sgn}\ x)y_{1\pm}(x),
\end{equation}
$\tau=\frac{i}{c}\sqrt{\frac{c^4}{4}+1}$, and
$e^{it}:=\left(\frac{c^2}{2}-i\right)\left(\sqrt{\frac{c^4}{4}+1}\right)^{-1}$,
see, e.g., \cite{AL}.

Using the definition of $S_\pm$ and (\ref{ura1}) we obtain
\begin{equation}\label{aaa19}
e_{++}=y_{1+}, \quad e_{+-}=y_{2+}, \quad e_{-+}=y_{1-}, \quad e_{--}=y_{2-}.
\end{equation}

The adjoint operator
$$
S^*=-ic\frac{d}{dx}\otimes{\sigma_1}+\frac{c^2}{2}\otimes\sigma_3
$$
is defined on the domain
$\mathcal{D}(S^*)={W}_2^1(\mathbb{R}\setminus\{0\})\otimes{\mathbb{C}}^2$
and an arbitrary $J$-self-adjoint extension $A_U\in\Sigma_J$
is the restriction of $S^*$ onto
${\mathcal D}(A_{U})={\mathcal{D}}(S)\dot{+}M$,
where $M$ is defined by  (\ref{e6}) and (\ref{aaa5}) with
$e_{\pm\pm}$ determined by (\ref{aaa19}).

It is easy to see that the fundamental symmetry $R=(\textsf{sgn}\ x)I$ in
$L_2(\mathbb{R})\otimes{\mathbb{C}}^2$ also commutes with $S$ and
$J{R}=-RJ$. Taking into account Remark \ref{did234} we establish the existence
of $J$-self-adjoint extensions of $S$ with empty resolvent set.

A routine calculation with the use of Corollary \ref{es200} gives that $A_U\in\Sigma_J$ has empty resolvent set
if and only if $A_U(=A_\gamma)$ is the restriction of $S^*$ onto the set ${\mathcal
D}(A_\gamma)=$
$$
\left\{\,y\in{W}_2^1(\mathbb{R}\setminus\{0\})\otimes{\mathbb{C}}^2
\ | \
\begin{array}{c}
\Lambda_\gamma[y(0+)+y(0-)]=y(0+)-y(0-) \vspace{2mm} \\
y'(0+)+y'(0-)=\Lambda_\gamma[y'(0+)-y'(0-)]
 \end{array} \right\},
$$
where $\Lambda_\gamma=\left(\begin{array}{cc}
e^{i\gamma} & 0 \\
0 & e^{-i\gamma}
\end{array}\right)$.

\noindent \textbf{Acknowledgements.}
The first author (S.K.) expresses his gratitude to the DFG
(project TR 903/11-1) for the support and the Fakult\"{a}t f\"{u}r Mathematik und Naturwissenschaften of Technische Universit\"{a}t Ilmenau for the warm hospitality.

\end{document}